\documentclass[referee,useAMS,usegraphicx]{mn2e}
\usepackage{amssymb}
\usepackage{graphicx}
\usepackage{color}

\title[Lithium in the $\alpha$ Per Cluster]
{Lithium Abundances in the $\alpha$ Per Cluster}

\author[Suchitra C. Balachandran, Sushma V. Mallik, and David L. Lambert]
{Suchitra C. Balachandran$^{1}$, Sushma V. Mallik$^{2}$\thanks{E-mail:
sgvmlk@iiap.res.in},
David L. Lambert$^{3}$\\
$^{1}$Astronomy Department, University of Maryland, College Park, MD
20742-2421, USA; suchitra.balachandran@gmail.com
\\
$^{2}$Indian Institute of Astrophysics, Bangalore 560034, India;
sgvmlk@iiap.res.in
\\
$^{3}$W. J. McDonald Observatory. The University of Texas at
Austin. 1 University Station, C1400.\\
Austin, TX 78712$-$0259, USA;
dll@astro.as.utexas.edu
\\}
\begin{document}
\date{}
\pagerange{\pageref{firstpage}--\pageref{lastpage}} \pubyear{2007}
\maketitle
\label{firstpage}
\begin{abstract}
Lithium abundances are presented and discussed for 70 members of the
50 Myr old open cluster $\alpha$ Per.
More than half
of the abundances are from new high-resolution spectra. The Li abundance
in the F-type stars is equal to its presumed initial abundance confirming
previous suggestions that pre-main sequence depletion is ineffective for these
stars. Intrinsic star-to-star scatter in Li abundance among these stars  is
comparable to the measurement uncertainties. There is marginal evidence that the
stars of high projected rotational velocity ({\it v} sin {\it i}) follow a different
abundance vs temperature trend to the slow rotators. For stars cooler than
about 5500 K, the Li abundance declines steeply with decreasing temperature
and there develops a star-to-star scatter in the Li abundance. This scatter
is shown to resemble the well documented scatter seen in the 70 Myr
old Pleiades cluster.  The scatter appears to be far less pronounced in the
30 Myr clusters which have been studied for Li abundance. 
\end{abstract}

\section{Introduction}

Abundance measurements of lithium in stellar atmospheres have long been an
active pursuit for observers and theoreticians alike. Much of the activity
is directed at understanding the depletion of the atmospheric lithium
from its abundance, often an inferred quantity, acquired
at  birth.
Open clusters serve as astrophysical laboratories in which to
investigate the internal depletion of lithium because
a given cluster provides a close approximation to a sample of coeval stars of a
common age and
initial composition including that of lithium
but spanning a range
of masses and  other properties such as rotation. And, crucially,
 the suite of clusters
spans a large range of ages for a small range in
composition. Three principal episodes of
lithium depletion are recognized: (i) depletion through destruction
of lithium at the base of the convective envelope of the pre-main sequence
star, (ii) continued depletion by destruction in the main sequence
phase, and (iii) depletion by a combination of diffusion and destruction
in F-type  main sequence stars in the narrow effective temperature
range of about 6400--6900 K (the so-called Li-dip). Sestito \& Randich
(2005) assemble Li abundance data for 20 open clusters with ages from
5 Myr to 8 Gyr to confront theories for Li depletion with
observations.
    
The cluster $\alpha$ Per was among the sample of 20 clusters  with
Li observations drawn from Balachandran, Lambert, \&
Stauffer (1988, 1996 -- hereafter BLS)
and Randich et al. (1998). In this paper, we obtain and analyze high-resolution
spectra from which Li abundances are obtained for about 50 stars. When the
BLS and this new sample are combined in a uniform manner and a reconsideration made of
the cluster membership of the stars, Li abundances are provided for
70 cluster  members.
                                                        
Observations of Li in $\alpha$ Per were  made initially by Boesgaard et al.
(1988) who analyzed high-resolution spectra of six F-type stars to show
that the Li-dip (Boesgaard \& Tripicco 1986) has not yet developed in this
young (age of about 50 Myr) cluster.
Our principal goal was not to define the run of Li abundance
along the main sequence because that is already well known  for
F, G, and K-type stars
(Boesgaard et al. 1988; BLS;  Randich et al. 1998) and M-type
stars  (Garc\'{i}a L\'{o}pez et al. 1994;
Zapatero Osorio et al. 1996). Rather we sought to determine if
the Li abundance at a given effective temperature has an intrinsic
scatter.
                                                                                                                             
Such a star-to-star variation in  apparent Li abundances has been
reported for the Pleiades, a cluster only slightly older than
$\alpha$ Per (Butler et al. 1987; Soderblom et al. 1993; King et al. 2000).
This variation appears for
stars with effective temperatures less than about 5300 K and
extends to the useful limit of the sample at about 4000 K. The
peak-to-peak variation is about 1.5 dex in apparent Li
abundance. Stars with the stronger Li\,{\sc i} 6707 \AA\
feature at a given temperature have higher projected
rotational velocities ({\it v} sin {\it i}).
The debate is ongoing as to whether the variation in Li\,{\sc i}
line strength in the Pleiades and other clusters
reflects a real abundance difference or
differences in
atmospheric structure not modelled by classical atmospheres.
                                                                                                                     
Randich et al. (1998) studied Li abundances in 18 very active, X-ray
selected members of $\alpha$ Per enlarging the original sample of BLS
in the 5500 $-$ 3900 K range of effective temperature. Randich et al. (1988) suggested 
that, at T$_{eff}$ $\le$ 5300K, there was indeed a significant dispersion in Li 
abundances in stars at the same temperature.  They further suggested that
rapid rotators had more Li and exhibited a smaller dispersion than slow rotators 
at the same T$_{eff}$. 
They inferred from these observations a likely
relationship between Li, chromospheric activity and the rotational history of stars.

 Examination of BLS's lithium observations as reanalyzed by
Randich et al. (1998) led Xiong \& Deng (2005)    to suggest
that star-to-star variations were also present among $\alpha$ Per
members at effective temperatures of about 4700 K and
that the variations
primarily arose from atmospheric effects and not a real abundance
variation. With our
larger sample of cluster members, we reexamine the question of
star-to-star variation in lithium abundance.
                                                                                                                             
In Section 2, we discuss selection of the newly observed stars. In Section 3,
we describe the new high-resolution spectra
of $\alpha$ Per
stars. Section 4 presents the stellar parameters with emphasis on
the effective temperature.
The abundance analysis is introduced in
Section 5. The run of Li abundance with effective temperature and the
star-to-star variations are discussed in Section 6. The paper concludes
with general remarks in Section 7.

\section{Construction of the final sample}
                                                                                                                             
In referring to members of the cluster, we follow the convention of `WEBDA', a website
devoted to stellar clusters\footnote{
http://www.univie.ac.at/webda/}. Heckmann et al. (1956) and Heckmann \& L\"{u}beck
(1958)
 introduced a numbering
scheme preceded by the letters `He'. Later, Stauffer et al. (1985, 1989a) and Prosser
(1992)
identified fainter stars with the letters `Ap'. In WEBDA, the label He is replaced
by \#, thus He 12 becomes \#12.  In the case of the Ap stars, the numbering is
increased by
1500 and Ap replaced by \#, thus Ap 79 becomes \#1579.
                                                                                                                             
When the observations (see below) made for this paper are combined with
those reported by Balachandran et al. (1988, 1996), we have spectra
for
86 stars. Since our primary goal  is to determine whether the
run of Li abundance down the main sequence of the cluster
exhibits scatter at a given effective temperature, it is vital
to sort cleanly the cluster members from the non-members and also
to separate out suitable from unsuitable
(i.e., doubled-lined spectroscopic binaries) members.

As long recognized, clean separation of members from non-members is not an easy
task for $\alpha$ Per because it is at low Galactic latitude and has a small
relative proper motion.
In making the separation, we have called upon a variety of publications
that have previously attempted the task.
The primary source and the one used in selecting stars for
observation was the seminal study of the cluster by Prosser (1992).
He considered a variety of membership indicators among which the
primary ones were proper motions and radial velocities of the
stars.

Makarov (2006) reanalysed the cluster's proper motions using astrometry
and photometry from the Tycho-2 Catalogue and the Second USNO CCD
Astrometric Catalog (UCAC2). Makarov's table of `High-Fidelity' members
lists 139 stars with a V magnitude brighter than about  11.5; no table of
non-members is provided. Of the stars in our sample with V$<$ 11.5 
and designated as members according to Prosser, all but ten
are among Makarov's high-fidelity members.  Four of the ten  stars
not listed as members by Makarov are categorized as non-members by Mermilliod
et al. (2008) (see below).  
It is unclear whether Makarov studied all stars from Prosser's list 
with V $<$ 11.5,
and therefore the absence of a star in Makarov's table is not necessarily  
an indication that it is not a member.  We have included the remaining six 
stars in our sample in Table 1.

Mermilliod et al. (2008) undertook a radial velocity program to
check for spectroscopic binaries in the cluster. Their criteria
for membership were threefold: proper motions (from  UCAC2), radial
velocity and location in the color-magnitude diagram. These
criteria were applied independently of Prosser's and Makarov's
efforts at membership determination.
Fifty four of our 86 stars were in Mermilliod et al.'s program.
Of these only four were identified as non-members in that
program: \# 143, 573, 1100, and 1181, with the first two shown to
be spectroscopic binaries.  We adopt Mermilliod  et al.'s
view that this quartet are non-members and list these stars
in Table 2.
Some  members were shown to be
spectroscopic binaries.  Binaries  not yet shown to be
double-lined are included in the list of 70 members and identified
in the final column in Table 1; all seven fall near the main sequence
locus in a color-magnitude diagram suggesting the secondary star contributes
very little to the composite spectrum.
                                                                                                                             
Patience et al. (2002) report on an imaging search for close binaries
among known cluster members; these authors made no independent determinations of
membership. A large fraction of our stars was examined
by Patience et al. with the majority reported not to have a companion
that would have contributed to our spectrum which we have
assumed is that of a single star. Four stars were excluded
as unsuitable for analysis on the basis of the reported imaging; these
have companions separated by less than 0.5 arc seconds and rather
similar masses. The stars are \# 696 (also known as \#1538), 935, 1541 and 1598.

In summary, 70 of the 86 stars are considered to be cluster
members (Table 1). Information provided in Table 1 is as follows: the WEBDA \# is in
column 1,
the adopted stellar parameters are in columns 2, 3, and 4. The projected rotational
velocity ({\it v} sin {\it i}) in column 7 is primarily taken from Prosser (1992). The
equivalent
width of the Li\,{\sc i} 6707 \AA\ feature is given for stars with low {\it v} sin {\it i} in
column 8 and the derived Li abundance is in column 9. Columns 10, 11, and 12
summarize the membership status of the star as given by
Prosser (1992), Makarov (2006) and Mermilliod et al. (2008). The final column
identifies the
seven stars that are spectroscopic binaries.
These seven are members and, it is assumed, that the secondary star is too faint to
contribute to the spectrum. They are therefore included with the single stars in
Table 1 and in our analyses. 

Sixteen stars, originally classified as members by
Prosser (1992), have subsequently been identified as non-members, single-lined or 
double-lined spectroscopic binaires, or close double stars.
The nature of these stars and the source of the revised information is listed in
Table 2 which has the same format as Table 1.   The stars are not rejected outright 
from our sample.  Rather, temperatures, rotational
velocities and Li and Fe abundances were determined as for the members and the results
are discussed with caveats and questions in Sections 6.2 and 6.3.

Our sample of 70 certain members and 16 stars possibly of questionable status represents the
largest selection to date for which lithium abundances are available in $\alpha$ Per.
                                                                                                                             
\begin{table*}
\caption{Cluster members, stellar parameters and lithium abundance}
\begin{tabular}{ccccccccccccc}
\hline
Star & {\it $T_{\rm eff}(V-K)$} & $\log g$ & $\xi_t$ & {\it $T_{\rm eff}(spec)$} & {\it $T_{\rm eff}(\beta)$} & {\it v} sin {\it i} & $W_\lambda$(Li)$^c$ &
$\log$ N(Li) &\multicolumn{3}{c}{ Membership$^a$} & Notes$^b$ \\
\# & (K) & cm s$^{-2}$ & km s$^{-1}$ & (K) & (K) & km s$^{-1}$ & (m\AA)  &   & Pr & Ma & Me & \\
\hline\hline 
  12 & 6663 & 4.5 & 1.5 &   & 6953 & 49 & 101 & 3.27 & Y & Y & Y & SB$?$  \\
  56 & 5703 & 4.5 & 0.8 & 5600 &   & 7 & 69 & 2.35 & Y & Y & $\dots$ & \\
  92 & 6683 & 4.5 & 1.5 &   &   & 23 & 109 & 3.31 & Y & $\dots$ & $\dots$ & \\
  93 & 5764 & 4.5 & 1.5 &   & 5880 & 25 & 119 & 2.66 & Y & $\dots$ & $\dots$ & \\
  94 & 5703 & 4.5 & 1.5 &   &   & 65 & 225 & 2.94 & Y & Y & $\dots$ & \\
 135 & 6903 & 4.5 & 1.5 &   & 6714 & 16 & 70 & 3.21 & Y & Y & Y & \\
 174 & 4928 & 4.5 & 1.5 & 5000 & 5319 & 12 & 196 & 2.25 & Y & Y & Y & \\
 270 & 6742 & 4.5 & 1.5 &   & 6491 & 33 & 96 & 3.24 & Y & Y & Y & SB \\
 299 & 6036 & 4.5 & 1.3 & 6200 &   & 15 & 106 & 2.99 & Y & Y & Y & \\
 309 & 6903 & 4.5 & 1.5 &   & 6448 & 65 & 62 & 3.15 & Y & Y & $\dots$ & \\
 334 & 6511 & 4.5 & 1.7 & 6400 & 7040 & 19 & 105 & 3.23 & Y & Y & Y & \\
 338 & 6606 & 4.5 & 1.5 &   &   & 56 & 111 & 3.25 & Y & Y & Y & \\
 350 & 5673 & 4.5 & 1.5 &   & 5893 & 42 & 215 & 2.91 & Y & Y & Y & \\
 361 & 7113 & 4.5 & 1.5 &   & 6740 & 30 & 59 & 3.18 & Y & Y & Y & \\
 421 & 6761 & 4.5 & 1.5 &   & 6935 & 90 & 58 & 3.05 & Y & Y & $\dots$ & \\
 490 & 7005 & 4.5 & 1.5 &   & 6821 & 17 & 76 & 3.31 & Y & Y & Y & \\
 520 & 5405 & 4.5 & 1.5 &   & 5468 & 91 & 317 & 2.87 & Y & $\dots$ & $\dots$ & \\
 588 & 6205 & 4.5 & 1.5 &   & 6532 & 120 & 76 & 2.75 & Y & Y & $\dots$ & \\
 621 & 6862 & 4.5 & 1.5 &   & 6613 & 28 & 76 & 3.25 & Y & Y & Y & \\
 632 & 7007 & 4.5 & 1.5 &   & 6632 & 160 & 76 & 3.31 & Y & Y & $\dots$ & \\ 
 660 & 6310 & 4.5 & 1.5 &   &   & 38 & 90 & 2.92 & Y$?$ & Y & Y & \\
 709 & 5873 & 4.5 & 1.5 &   &   & 59 & 187 & 3.01 & Y & Y & $\dots$ & \\
 750 & 6437 & 4.5 & 1.5 &   &   & 26 & 170 & 3.34 & Y & Y & Y & \\
 767 & 6222 & 4.5 & 1.3 & 6100 &   & 10 & 139 & 3.27 & Y & Y & Y & \\
 799 & 7244 & 4.5 & 1.5 &   & 6622 & 49 & 70 & 3.30 & Y & Y & $\dots$ & \\
 828 & 5503 & 4.5 & 1.5 &   &   & 12 & 157 & 2.68 & Y & Y & Y & \\
 833 & 6702 & 4.5 & 1.5 &   & 6491 & 27 & 144 & 3.46 & Y & $\dots$ & Y & \\
 841 & 6530 & 4.5 & 1.5 &   &   & 65 & 86 & 3.04 & Y & Y & $\dots$ & \\
 917 & 6003 & 4.5 & 1.5 &   & 5841 & 40 & 191 & 3.11 & Y & $\dots$ & Y & \\
 968 & 6474 & 4.5 & 1.5 &   &   & 30 & 200 & 3.51 & Y & Y & Y & \\
 972 & 6455 & 4.5 & 1.5 &   & 6491 & 87 & 93 & 3.08 & Y & Y & $\dots$ & \\
1086 & 5749 & 4.3 & 1.4 & 5900 & 6122 & 12 & 151 & 2.96 & Y & Y & Y & \\
1101 & 5540 & 4.5 & 1.5 &   & 5387 & 35 & 377 & 3.11 & Y & Y & $\dots$ & \\
1180 & 6761 & 4.5 & 1.5 &   & 6522 & 45 & 99 & 3.32 & Y & Y & Y & \\
1185 & 5718 & 4.5 & 1.2 & 6000 & 5669 & 7 & 132 & 2.91 & Y & Y & Y & SB \\
1514 & 5503 & 4.3 & 1.0 & 5400 &   & 8 & 187 & 2.94 & Y & $\dots$ & Y & \\
1519 & 5417 & 4.5 & 1.5 &   &   & 50 & 293 & 2.81 & Y & Y & $\dots$ & \\
1525 & 5187 & 4.3 & 1.6 & 5300 &   & 12 & 193 & 2.64 & Y & Y & Y & \\
1528 & 4757 & 4.3 & 1.5 & 4900 &   & 12 & 70 & 1.32 & Y & $\dots$ & Y & \\
1532 & 6419 & 4.5 & 1.5 &   &   & 65 & 148 & 3.26 & Y & Y & Y & SB \\
1533 & 4889 & 4.5 & 1.5 &   &   & $<$ 10 & 143 & 1.79 & Y & $\dots$ & Y & \\
1537 & 5008 & 4.5 & 1.7 & 5200 &   & 20 & 83 & 1.77 & Y & $\dots$ & Y & \\
1543 & 4695 & 4.5 & 1.5 &   &   & 72 & 539 & 2.50 & Y & $\dots$ & $\dots$ & \\
1551 & 6400 & 4.5 & 1.5 &   &   & 65 & 92 & 3.00 & Y & Y & $\dots$ & \\
1556 & 4757 & 4.5 & 1.5 &   &   & 110 & 299 & 2.11 & Y & $\dots$ & $\dots$ & \\
1565 & 4832 & 4.5 & 0.9 & 4800 &   & 10 & 61 & 1.27 & Y & $\dots$ & Y & \\
1570 & 4908 & 4.3 & 1.9 & 5300 &   & 7 & 169 & 2.26 & Y & $\dots$ & Y & \\
1572 & 5018 & 4.5 & 1.4 & 5100 &   & 10 & 110 & 2.03 & Y & $\dots$ & Y & \\
1575 & 4072 & 3.8 & 2.2 & 4900 &   & 11 & 43 & 0.17 & Y & $\dots$ & Y & SB  \\
1578 & 4804 & 4.3 & 2.0 & 5200 &   & 13 & 162 & 1.94 & Y & $\dots$ & $\dots$ & \\
1589 & 5381 & 4.3 & 1.0 & 5900 &   & 8 & 77 & 2.24 & Y & $\dots$ & Y$?$ & \\
1590 & 5970 & 4.5 & 1.5 &   &   & 12 & 148 & 3.08 & Y & Y & Y & \\
1591 & 4822 & 4.5 & 1.5 &   &   & 25 & 232 & 2.22 & Y & $\dots$ & $\dots$ & \\
1593 & 4785 & 4.5 & 1.5 &   &   & 75 & 365 & 2.26 & Y & $\dots$ & $\dots$ & \\
1597 & 5232 & 4.5 & 1.5 &   &   & 10 & 196 & 2.70 & Y & Y & Y & SB  \\
1600 & 4315 & 4.5 & 1.5 &   &   & 205 & 69 & 0.55 & Y & $\dots$ & $\dots$ & \\
1601 & 4286 & 4.5 & 1.5 &   &   & $<$ 10 & 60 & 0.48 & Y & $\dots$ & Y & \\
1604 & 5598 & 3.8 & 1.1 & 5900 &   & 8 & 45 & 2.06 & Y & $\dots$ & Y & \\
1606 & 4889 & 4.0 & 2.2 &   &   & 8 & 158 & 1.90 & Y & $\dots$ & Y & \\
1607 & 4948 & 4.5 & 1.5 &   &   & 9 & 129 & 2.00 & Y & $\dots$ & Y & \\
1610 & 5187 & 4.3 & 1.5 & 5200 &   & 8 & 205 & 2.62 & Y & $\dots$ & Y & \\
\hline
\end{tabular}
\end{table*}
\begin{table*}
\begin{tabular}{ccccccccccccc}
\hline
Star & {\it $T_{\rm eff}(V-K)$} & $\log g$ & $\xi_t$ & {\it $T_{\rm eff}(spec)$} & {\it $T_{\rm eff}(\beta)$} & {\it v} sin {\it i} & $W_\lambda$(Li)$^c$ &
$\log$ N(Li) &\multicolumn{3}{c}{Membership$^a$} & Notes$^b$ \\
\# & (K) & cm s$^{-2}$ & km s$^{-1}$ & (K) & (K) & km s$^{-1}$ & (m\AA)  &   & Pr & Ma & Me & \\
\hline\hline
1612 & 4405 & 4.5 & 1.5 &   &   & 13 & 10 & $-$0.65 & Y & $\dots$ & $\dots$ & \\
1614 & 4524 & 4.5 & 1.5 &   &   & 12 & 120 & 1.30 & Y & $\dots$ & Y & \\
1617 & 4712 & 4.5 & 1.5 &   &   & 83 & 469 & 2.35 & Y & $\dots$ & $\dots$ & \\
1618 & 5133 & 4.5 & 1.5 &   &   & 160 & 307 & 2.51 & Y & $\dots$ & $\dots$ & \\
1621 & 5405 & 4.3 & 1.3 & 5500 &   & 10 & 159 & 2.67 & Y & $\dots$ & Y$?$ & \\
1669 & 4557 & 4.0 & 1.6 & 4800 &   & 8 & 54 & 0.86 & Y & $\dots$ & Y & \\
1697 & 4767 & 4.3 & 1.7 & 5000 &   & 10 & 78 & 1.49 & Y & $\dots$ & Y & \\
1731 & 4228 & 4.5 & 0.8 & 4500 &   & 25 & 56 & 0.38 & Y & $\dots$ & $\dots$ & \\
1735 & 4651 & 4.3 & 2.2 & 4900 &   & 11 & 24 & 0.34 & Y & $\dots$ & Y & \\
\hline
\end{tabular}

\noindent
$^a$ Pr = Prosser (1992), Ma = Makarov (2006), Me = Mermilliod et al.
(2008)
$^b$ All SB and SB$?$ designations from Mermilliod et al. (2008)
except for
\#727 from Prosser (1992)
$^c$ For stars with {\it v} sin {\it i} $>$ 25 km s$^{-1}$, EQWs were not measured but derived from the 
Li abundance determined from spectrum synthesis.
\end{table*}

%
%
                                                                                                                                               
\begin{table*}
\caption{Non-members, binaries and doubles}
\begin{tabular}{ccccccccccccc}
\hline
Star & {\it $T_{\rm eff}(V-K)$} & $\log g$ & $\xi_t$ & {\it $T_{\rm eff}(sp)$} & {\it $T_{\rm eff}(\beta)$} & {\it v} sin {\it i} & $W_\lambda$(Li)$^c$ &
$\log$ N(Li) &\multicolumn{3}{c}{Membership$^a$} & Notes$^b$ \\
\# & (K) & cm s$^{-2}$ & km s$^{-1}$ & (K) & (K) & km s$^{-1}$ & (m\AA)  &   & Pr & Ma & Me & \\
\hline\hline
 143 & 5873 & 4.0 & 1.0 & 5700 & 6243 & 10 & 83 & 2.62 & Y & $\dots$ & N & SB1O \\
  &   &   &   &   &   &   &   &   &   &   &   & (Pr,Me) \\
 407 & 5937 &   &   &   &   & 28 & 29 & 2.01 & Y & $\dots$ & $\dots$ & \\
 573 & 6549 & 4.0 & 0.6 & 6600 & 6782 & 12 & $<$5 & 1.69 & Y & $\dots$ & N & SB \\
  &   &   &   &   &   &   &   &   &   &   &   & (Me) \\
 715 & 6903 &   &   &   & 6522 & 110 & 104 & 3.41: & Y & Y & $\dots$ & SB2$?$ \\
  &   &   &   &   &   &   &   &   &   &   &   & (Pr,Ma) \\
 848 & 6346 & 4.5 & 1.3 & 6500 &   & 16 & 95 & 3.15: & Y & Y & Y & SB2O \\
  &   &   &   &   &   &   &   &   &   &   &   & (Pr,Ma,Me) \\
 935 & 6119 &   &   &   &   & 56 & 176 & 3.13 & Y & Y & $\dots$ & Double \\
  &   &   &   &   &   &   &   &   &   &   &   & (Ma) \\
1100 & 5528 & 4.5 & 0.8 & 5800 &   & 8 & 59 & 2.06 & Y & $\dots$ & N &  \\
1181 & 6205 & 4.0 & 1.1 & 5700 & 6034 & 7 & 58 & 2.72 & Y & $\dots$ & N &  \\
1234 & 5658 & 4.5 & 1.6 & 6000 &   & 10 & 90 & 2.56: & Y & Y & Y & SB2 \\
  &   &   &   &   &   &   &   &   &   &   &   & (Me) \\
1538 & 5613 & 4.5 & 1.5 & 5700 &   & 10 & 193 & 2.95 & Y & Y & Y & Double \\
1541 & 5288 & 4.3 & 1.5 & 5400 &   & 8 & 200 & 2.76 & Y & $\dots$ & Y & Double \\
1598 & 4938 &   &   &   &   & 10 & 199 & 2.30 & Y & $\dots$ & Y & Double \\
1602 & 5381 & 4.3 & 1.0 & 5600 &   & 11 & 141 & 2.56: & Y & Y & Y & SB2 \\
  &   &   &   &   &   &   &   &   &   &   &   & (this work) \\
1625 & 5358 & 4.3 & 1.8 & 5700 &   & 48 & 108 & 2.33: & Y & $\dots$ & $\dots$ & SB2 \\
  &   &   &   &   &   &   &   &   &   &   &   & (this work) \\
1656 & 5311 & 4.3 & 0.8 & 5600 &   & 8 & 103 & 2.28: & Y & $\dots$ & Y & SB2 \\
  &   &   &   &   &   &   &   &   &   &   &   & (Me) \\
1713 & 5243 & 4.3 & 0.7 & 5500 &   & 5 & 18 & 1.04: & Y & $\dots$ & Y & SB2 \\
  &   &   &   &   &   &   &   &   &   &   &   & (Pr,Me) \\
\hline
\end{tabular}
\noindent
$^a$ Pr = Prosser (1992), Ma = Makarov (2006), Me = Mermilliod et al.
(2008)
                                                                                                                                               
\noindent
$^b$ Classifications as SB from various sources : Mermilliod et al.,
Makarov, Prosser
and
our observations. Double denotes a close binary reported by Patience et al. (2002).
\#407
is a
non-member according to Fresneau (1980) and unusually reddened (Trullols et al. 1989).

\noindent
$^c$ For stars with {\it v} sin {\it i} $>$ 25 km s$^{-1}$, EQWs were not measured but derived from the
Li abundance determined from spectrum synthesis. 
\end{table*}
                                                                                                                             
\section{Observations}
                                                                                                                             
High-resolution spectra were obtained between 1992 and 1994. Observations were
made during December 1992 and
November 1993 for 30 stars at the 2.7m telescope at the W.J. McDonald
Observatory with the Robert G. Tull cross-dispersed echelle spectrograph (Tull et
al. 1995) at
a resolving power of about 60,000 with exposure times chosen to provide a
S/N ratio of 100 or higher.
In January 1994,
observations were carried out for 21 stars at the 4m telescope at KPNO with the
Cass\'{e}grain  echelle spectrograph, the red long-focus camera and the Tex 2048 x 2048 
CCD chip to give a 2-pixel resolution of 0.16 \AA\ (R $\sim$ 40,000).  
Integration times were chosen to provide a S/N ratio close to 150 for most stars and even
higher in a few cases.

\noindent
Data reduction was carried out following standard IRAF procedures.\footnote{IRAF
     is distributed by the National Optical Astronomy Observatories,
    which are operated by the Association of Universities for Research
    in Astronomy, Inc., under cooperative agreement with the National
    Science Foundation.}
The frames were trimmed and overscan corrected.  Bias frames were combined and
subtracted from the raw spectrum.  The spectrum was divided by the
normalized flat field image to account for the pixel to pixel sensitivity
difference of the detector and then corrected for scattered light. No sky
subtraction was done as the sky signal was negligible in all cases.  

Nineteen 
and twenty-four echelle orders were
extracted respectively from the McDonald and the KPNO data.  The wavelength
scale for all the orders was derived using the Thorium-Argon spectrum. The
wavelength calibrated spectrum was then normalized to a continuum of one.
                                                                                                                             
The measured equivalent widths (EQW)
of the Li\,{\sc i} line at 6707.8 \AA\   are given in Tables 1 and 2
for stars with low projected rotational velocities ({\it v} sin {\it i} $<$ 25 km s$^{-1}$). 
The EQWs include the contribution of the Fe I blend at 6707.435 \AA.  The
contribution of the Fe I blend was removed by the program MOOG (Sneden 1973) during the derivation
of the Li abundance.  
In the slow rotators ({\it v} sin {\it i} $<$ 25 km s$^{-1}$), the uncertainty in 
the  EQW, largely caused by the
placement of the continuum, was 2-3 m\AA\  at $\sim$
15 m\AA, 6 m\AA\ at $\sim$ 130 m\AA, and 10 m\AA\ at $\sim$ 200
m\AA.  In the spectra of the more rapidly rotating stars in which lines were still measurable,
the EQW uncertainty was estimated to be as large as $\sim$ 15 m\AA. EQWs were not measured
in these stars, rather Li abundances were determined by spectral synthesis, 
and the EQWs listed in Tables 1 and 2 for stars with {\it v} sin {\it i} $>$ 25 km s$^{-1}$ 
were calculated from the
derived abundance using MOOG (Sneden 1973).

The analysis of the 6707 \AA\  line was done for all the stars with
spectrum synthesis fits to the observed spectrum. For the 36 slowly
rotating stars for which spectroscopic analysis was possible, the Li
abundance was determined in addition from the Li EQW. The match between
the two measurements of Li abundance was in excellent agreement.

\section{Stellar Parameters}
                                                                                                                             
The Li abundance determined from the 6707 \AA\ feature, is primarily sensitive to
the adopted effective temperature $T_{\rm eff}$.
An error of $\pm$ 200 K in $T_{\rm eff}$, results in an uncertainty in $\log$ N(Li)
between $\pm$ 0.28 to $\pm$ 0.14 over the 4500 K to 6500 K temperature range.
Thus, we devoted considerable effort to a determination of
$T_{\rm eff}$. The Li abundance is quite insensitive
to the adopted surface gravity; a variation in $\log g$ of $\pm$0.5 dex results in
a change in Li abundance by less than $\pm0.02$ dex.
The adopted microturbulence has a small
influence on Li when the 6707 \AA\ feature is strong.
                                                                                                                             
\subsection{Effective Temperature}
                                                                                                                             
The effective temperature is derived from photometry, primarily the $(V-K)$ index,
and checked by use of the Str\"{o}mgren $\beta$ index and spectroscopy.
                                                                                                                             
\subsubsection{Photometry}
                                                                                                                             
Our principal photometric indicator of effective temperature is the (V-K) colour index
 which is available for all the stars.
All of
the observed stars have a K$_{\rm s}$ magnitude in the 2MASS
catalogue.\footnote{http://www.ipac.caltech.edu/2mass/releases/allsky/}
The K$_{\rm s}$ were transformed to Johnson K magnitudes by the
Koornneef transformations (Carpenter 2001). The V magnitudes were taken
from Prosser (1992).

The literature contains various estimates of the reddening affecting the cluster.
Several
authors refer to a variable reddening across the cluster. Cluster members
are slightly reddened but there is  little solid evidence that the reddening is
significantly
different from star-to star. BLS adopted E(B$-$V) = 0.08 (Mitchell 1960) for all their
stars and remarked that Crawford \& Barnes (1974) suggested a range from 0.04 to 0.21.
BLS note that the extremities of the range correspond to effective temperatures
lower by 100 K and hotter by 450 K. Thus, the larger reddenings are a concern in the
search for the origin of  a scatter in Li abundances.
                                                                                                   
Inspection of Crawford \& Barnes (1974) shows, however,  little evidence for a
variation in
reddening. In their Table III, they list measurements of $E(b-y)$ for 21 F-type stars.
Fifteen stars are members and six are non-members according to Makarov (2006).
(Twelve of the
15 members are in Table 1.)  The mean $E(b-y)$ for the 15 is 0.054$\pm0.017$ with
extremes of 0.032 and 0.091.  In contrast, larger reddening is seen among non-members with
the six non-members exhibiting a range in $E(b-y)$ from
0.023 to 0.148. Crawford \& Barnes did note that  18 A-type and 31 B-type cluster members gave
higher and similar reddening: fourteen A-type stars, members according to Makarov (2006), give
a mean $E(b-y)$ = 0.089$\pm0.036$ with extreme values of 0.038 and 0.139. The
factor of 1.7 between the mean values for the F- and A-type stars points to
an issue with the calibration.  Possibly, the larger standard error
in the A stars may indicate non-uniform colors resulting perhaps from diffusion
or metallicity effects.
 Trullols et al. (1989) provide $E(b-y)$ for 12 F star members:
the mean $E(b-y)$ is  0.065$\pm$0.012 where the
standard error again indicates little if no variation from star-to-star. Pe\~{n}a \&
Sareyan (2006) provide Str\"{o}mgren photometry for cluster stars from a combination
of their own measurements and published values and obtained a mean reddening from
169 stars of $E(b-y)$ = 0.073$\pm0.038$.  However, for the 15 F stars in common, their
reddening ($E(b-y)$ = 0.086$\pm0.029$) differs from that of
Crawford \& Barnes, suggesting calibration differences.
                                                                                                   
\noindent
Crawford (1975) used the available uvbyH$\beta$ photometry for bright stars and
cluster members to calibrate H$\beta$ in terms of intrinsic colour $(b-y)$ and other
indices, applicable for $\beta$ between 2.59 and 2.72. 
Using this calibration, we made fresh estimates of
$E(b-y)$ for 40 stars with the available Stromgren photometry
(from WEBDA which essentially includes the observations of Crawford \& Barnes (1974)
and Trullols et al. (1989)) in the above range of $\beta$ taking care that the sample
contained no binaries or possible non-members.  We found that $E(b-y)$ ranged from 
0.02 to 0.12, very similar to previous studies, with an average $E(b-y)$ of 0.075 and a 
standard deviation of the measurements (standard error) of $\pm$0.04.  For the normal 
interstellar reddening law, $A_V$ = 0.32, this translates to  $E(V-K)$ = 0.284 $\pm$ 0.15. 
The standard error may be used as one estimate of the reddening uncertainty for each star.

Prosser (1992) derived
$E(V-I)$ for about 75 M cluster dwarfs from a somewhat unusual process.
Low-dispersion spectra
provided  spectral types which with a color-spectra type relation gave the intrinsic
$(V-I)$ color
of a star.
Comparison of intrinsic and observed $(V-I)$ gave a star's reddening.
The mean $E(V-I)$ $\simeq 0.18$ corresponds to $E(b-y)$ $\simeq 0.08$  and $E(V-K)$
$\simeq 0.3$,
values consistent with other reddening measures from traditional techniques applied
to earlier
spectral types. Given that the reddenings are
described by Prosser as `preliminary values', one may attach little weight to his
suggestion
that the reddening is not uniform for cluster members.

The intrinsic $(V-K)$ colour was calculated assuming thus an average
interstellar reddening of $E(V-K)$ = 0.284.
The photometric calibrations
$(V-K)$ - $T_{\rm eff}$ of Alonso et al. (1996) given below were used to derive
the $(V-K)$ effective temperature $T_{\rm eff}(V-K)$
for all the stars.  This procedure was applied to all stars including those previously
analysed by BLS.  

\begin{equation}
\theta_{\rm eff} = 0.555 + 0.195(V-K) + 0.013(V-K)^2   
\end{equation}
for 0.4 $\le$ $(V-K)$ $\le$ 1.6 and

\begin{equation}
\theta_{\rm eff} = 0.566 + 0.217(V-K) - 0.003(V-K)^2   
\end{equation}
for 1.6 $\le$ $(V-K)$ $\le$ 4.1

\noindent
where $\theta_{\rm eff}$ = 5040/$T_{\rm eff}$.    We denote this photometric
temperature by $T_{\rm eff}(V-K)$  (see Table 1 for cluster members). 
The photometric error in K is 0.02 mag.\footnote{http://www.ipac.caltech.edu/2mass/releases/allsky/} 
and in V it is 
0.058 mag. (Prosser 1992), yielding a standard deviation in $(V-K)$ of 0.06 mag. 
An error in $(V-K)$ of 0.06 results in an error of 125K at 6600 K, 50K at 4500K and 75K at 5500K.
We caution that the presence of a cool companion may 
increase the $(V-K)$ color of the primary component of a binary and 
therefore the temperature derived for these stars (Table 2) may be
systematically low. We note here that the standard deviation
of 0.15 in $E(V-K)$ derived from the reddening calculation is substantially 
higher than the uncertainty in the photometric measurement
determined above and translates to temperature uncertainties of 325 K at 6600 K, 180 K at 
5500 K, and 125 K at 5500 K.   

Other colour indices might be considered as thermometers.
Several previous studies of $\alpha$ Per (and other clusters) have employed
$(B-V)$ for which measurements are available for all but five of our stars.
However, the V vs. $(B-V)$ colour-magnitude diagrams for young clusters are
fundamentally  different from the V vs. $(V-I)$ plots when overlaid by theoretical
isochrones. The best-fitting isochrone in the V vs. $(B-V)$ diagram
for young clusters such as $\alpha$ Per
follows the observed main sequence down to the late-K stars which
lie conspicuously to the left of the theoretical  main sequence; they are
consistently fainter and bluer.
This is
seen in the Pleiades and $\alpha$ Per  (S.V. Mallik, private communication).
Stauffer et al. (2003)
first pointed out this blue anomaly in the Pleiades and suggested it may be due to
a flux contribution (larger in B than in V) not represented by the model
atmospheres.
This so-called $(B-V)$ anomaly is predominant in the younger clusters
but is
absent in clusters as old as Praesepe: $\alpha$ Per is about 0.05 Gyr to Praesepe's
0.6 Gyr.
If this interpretation rather than a more deep-seated deficiency in
the models providing the isochrones is correct, the anomaly is
likely related to stellar surface activity that decays as stars age.
This has two obvious consequences for deriving and interpreting Li abundances.
First, $(B-V)$ may be a poorer temperature indicator for the cooler stars than
$(V-K)$;  the
colour-temperature calibration is likely dependent on a star's age but may also vary
from star-to-star with changes in stellar  activity. Then, these dependencies may
provide a star-to-star scatter in Li abundance.

\subsubsection{The Str\"{o}mgren $\beta$ index}

As a reddening-free index, the Str\"{o}mgren $\beta$ index is a useful
measure of effective temperature for stars hotter than about 5000 K.
Measures of $\beta$ are taken from Crawford \& Barnes (1974) and Trullols et al.
(1989). The $\beta$ vs $T_{\rm eff}$ calibration is taken from
Alonso et al. (1986 - see also Castelli \& Kurucz 2006).
$T_{\rm eff}(\beta)$ is listed in column 6 of Table 1 for 24 cluster members 
and the comparison 
$T_{\rm eff}(\beta)$ vs $T_{\rm eff}(V-K)$ is shown in
Figure 1. 
$T_{\rm eff}(\beta)$ is also listed for four stars of questionable status in Table 2
and included in Figure 1.  

While there is good agreement between $(V-K)$ and H$\beta$ temperatures at the 
cool end of the useful range of the $\beta$ index, there is a surprisingly larger scatter,
at the warm end, with temperature differences as large as 500K for the same star from the 
two calibrations.  The difference is also surprising because the two calibrations
come from a common paper.  We have no leads on whether this difference is due
to calibration issues or photometry errors but we surmise it is unlikely to be
caused by variable reddening as it is confined to a small temperature range.
The non-members and SB2s lie within the scatter defined by the members.
                                                                                                                             
\begin{figure}
\centering
\includegraphics[height=9cm,width=8cm,angle=90]{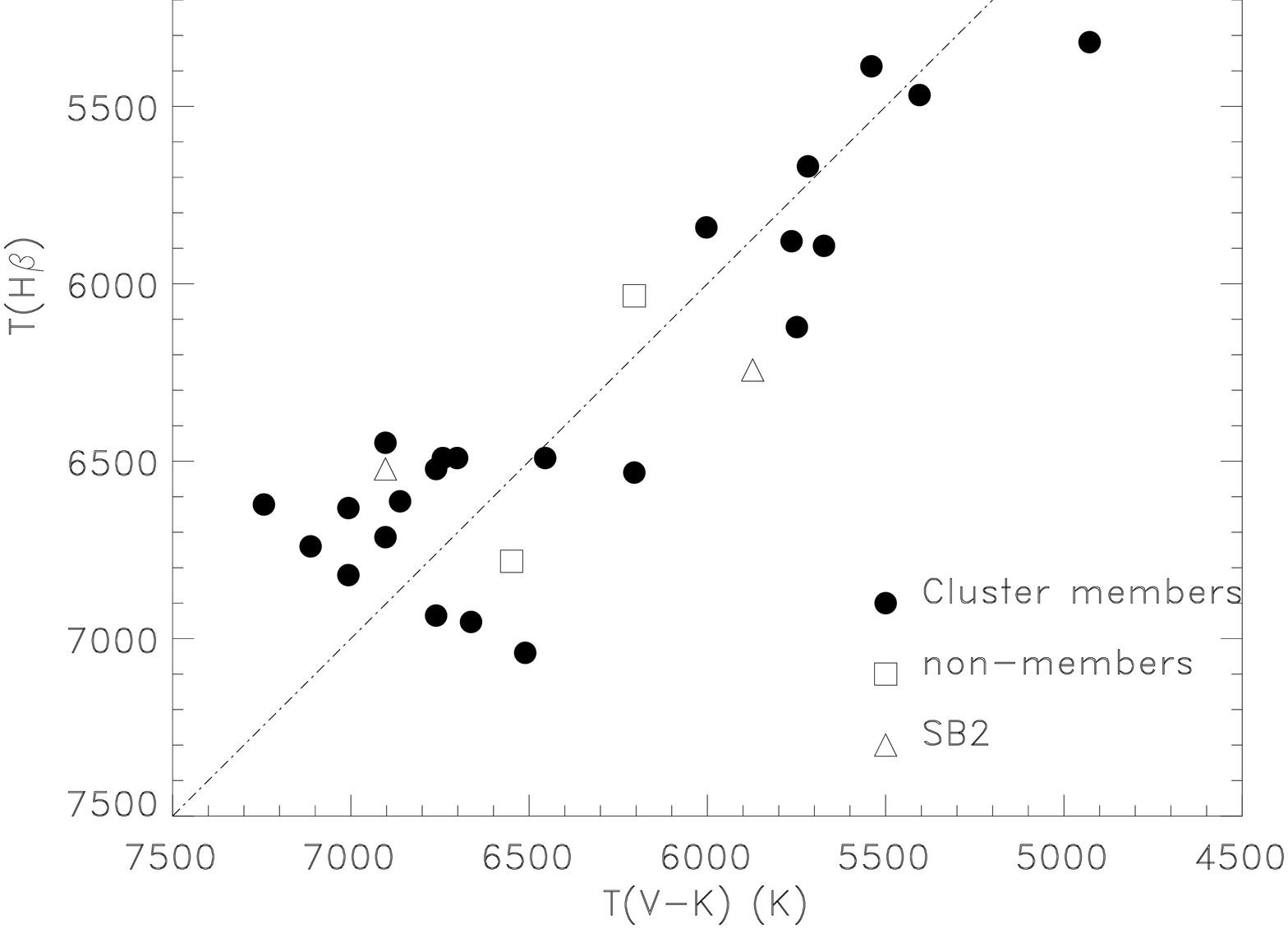}
\caption{Comparison of the effective temperatures derived from the
$(V-K)$ and $\beta$ indices. The line corresponds to perfect correspondence
between the two measurements. The symbols are described in the key.}
\end{figure}

\subsubsection{Spectroscopy}
                                                                                                                             
Our spectroscopic temperature is based on the usual condition that the
Fe\,{\sc i} lines in the observed spectra return the same Fe abundance independent
of a line's
lower excitation potential.
The McDonald spectra provide about 30 Fe\,{\sc i} lines spanning about
4 eV in the lower excitation potential. The KPNO spectra with their
greater wavelength coverage yield about 130 Fe\,{\sc i} lines. These numbers
pertain to slowly rotating stars  ({\it v} sin {\it i} $\le$ 20 km s$^{-1}$); more
rapidly rotating stars have broader lines that lead to blending and a difficulty in
measuring EQWs accurately, especially of weak lines.
The $T_{\rm eff}$ determination
has to be made simultaneously with that for the microturbulence $\xi_t$. For this
exercise in
determining $T_{\rm eff}$ and $\xi_t$, we used astrophysical 
$gf$-values for Fe\,{\sc i} 
and Fe\,{\sc ii} lines.
 The $gf$-values were determined using MOOG (Sneden 1973) with measurements of Fe\,{\sc i} and Fe\,{\sc ii}  
equivalent widths from the high resolution digital solar atlas (Delbouille et al. 1990), 
the Kurucz solar model atmosphere with no convective overshoot (Castelli, Gratton \& Kurucz 
1997) and requiring the lines to yield Fe/H=7.50 at $\xi_t$ = 0.8 km s$^{-1}$. 

\noindent
We  derived spectroscopic temperatures for 24 cluster members including eight
stars observed
by BLS for which we were able to retrieve their spectra (Table 1), and for 12 
binaries, doubles or non-members (Table 2).  The microturbulence and
spectroscopic temperatures are listed in columns 4 and 5 respectively.
A 100 K change
in effective temperature resulted in a significant non-zero slope of Fe I vs. lower
excitation potential to allow us to constrain temperatures to $\pm 100 K$.  
Errors in equivalent width measurement, gf-values and 
microturbulence would result in random and systematic temperature errors and we feel that
$\pm$ 200 K conservatively constrains the error in our derived $T_{\rm eff}(spec)$.  
The $\xi_t$ is determined to about $\pm$ 0.1 km s$^{-1}$.

A comparison of $T_{\rm eff}(V-K)$ and the
spectroscopic temperature $T_{\rm eff}(spec)$  is presented in Figure 2. 
On average, $T_{\rm eff}(V-K)$ is cooler than $T_{\rm eff}(spec)$ by about 250 K.
The temperature difference appears to vanish for the hotter stars, say
$T > 6000$ K. This level of agreement is consistent with the estimated
uncertainty from the analysis of the Fe\,{\sc i} lines and lends support to
the assertion that reddening is not very variable across the cluster.
The visual doubles, the SB2s and three of the four non-member stars lie within
the scatter defined by the cluster members.  Only \#1181 has a much cooler 
$T_{\rm eff}(spec)$ compared to
$T_{\rm eff}(V-K)$.

\begin{figure}
\centering
\includegraphics[height=9cm,width=8cm,angle=90]{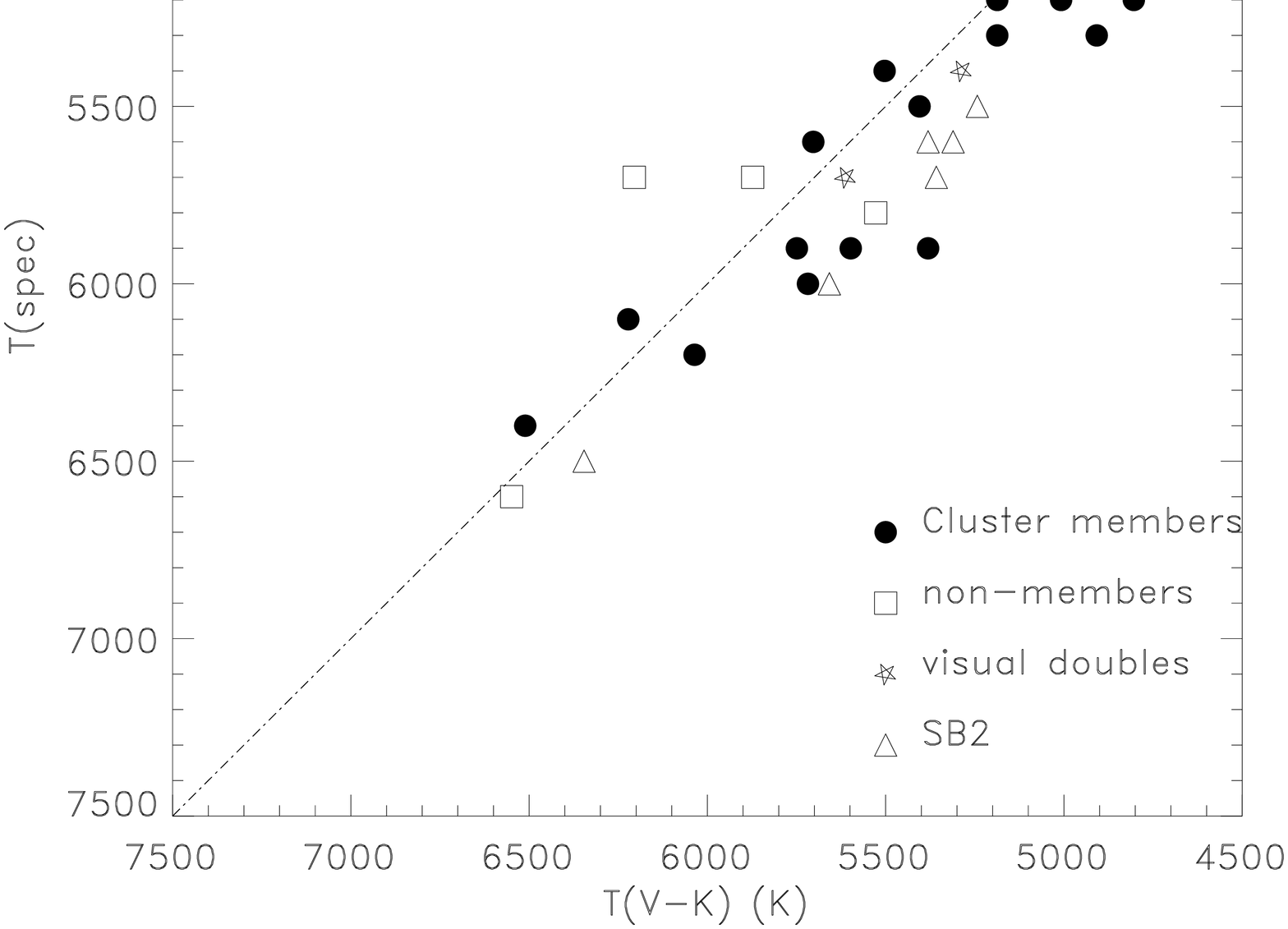}
\caption{Comparison of the effective temperatures derived from
the $(V-K)$ index and
the  Fe\,{\sc i} lines ($T_{\rm eff}(spec)$)
The line corresponds to perfect correspondence
between the two measurements. The symbols are described in the key.}
\end{figure}

\subsection{Surface gravity}
                                                                                                                             
With the inclusion of Fe\,{\sc ii} lines in the spectroscopic  analysis, it is
possible to
determine  the surface
gravity $\log g$. Spectra of  18 stars provide an adequate number of eight to ten
Fe\,{\sc ii} lines. A $\log g$ determination requires the same Fe abundance
from Fe\,{\sc i} and Fe\,{\sc ii} lines and this is possible to an accuracy of $\pm$
0.25  dex.
For the stars without a
spectroscopically
determined $\log g$, we adopt the $\log g$ = 4.5 for the $T_{\rm eff}$
determination from the Fe\,{\sc i} lines.
This value was adopted for all stars without a spectroscopic determination of
surface gravity.

\subsection{Microturbulence}
                                                                                                                             
The microturbulence is taken either from the analysis of the Fe\,{\sc i} lines or a
value of
1.5 km s$^{-1}$ was assumed.

\section{Abundance analysis}

\subsection {Lithium Abundances}
                                                                                                                             
The abundance analysis from which we extract the Li abundance takes
the standard form. 

Model atmospheres were generated in 100 K intervals in temperature and 0.1 dex intervals
in gravity
using the program ATLAS9, written and supplied by R. L. Kurucz.  Standard solar opacity
distibution functions were used with overshoot turned off (see Castelli, Gratton,
\& Kurucz 1997).  The appropriate model was chosen for each star according to the
stellar parameters listed in Tables 1 and 2.

The line analysis
program MOOG (Sneden 1973) was used to convert EQWs of the
6707 \AA\ Li\,{\sc i} resonance doublet to an abundance.  Throughout
the assumption of local thermodynamic equilibrium (LTE) is adopted.
The  $gf$-values and wavelengths  of the fine- and hyperfine-structure
components
of the Li\,{\sc i} feature were taken from Andersen, Gustafsson \& Lambert (1984).
The Li abundance was chosen by the best fit of a synthetic spectrum to a
region around the 6707 \AA\ feature with the line list adopted by BLS. It is
most unlikely there is any $^6$Li in stars where $^7$Li is even slightly depleted.
Therefore, $^6$Li was included in the line list only for stars with $log$ N(Li) 
$>$ 3.0.

Lithium abundances were computed for the model parameters listed
in Tables 1 and 2. 
 An error of $\pm$100K in $T_{\rm eff}$, $\pm$0.25 dex in $\log g$, 
$\pm$0.1 kms$^{-1}$ in $\xi_t$ and 
5 m\AA\ in Li I EQW results in Li abundance errors of $\pm$0.1, $\pm$0.01, $\pm$0.00 and $\pm$0.09
respectively. 
As these estimates show, the two principal sources of uncertainty
affecting the derived lithium abundances arise from the effective
temperature and the measured equivalent width. The effect of a 200 K
spread in effective temperature is shown in Figure 3 by the shaded
area at the bottom of the figure. 
At the lowest temperatures where the
stars of the same effective temperature can show lithium lines of quite
different strengths, the uncertainty in measurement of the equivalent
width may have a larger effect on the derived abundance than the
temperature uncertainty, especially for those few stars where the lithium
line is weak.  
The reddening uncertainty estimated from H$\beta$ increases the temperature error
over 200K only at the hottest temperatures; the uncertainty in $E(V-K)$ of 0.15 
translates to a temperature
uncertainty of 325 K at 6600 K.  This uncertainty would merely increase the Li abundance 
uncertainty at 6600 K to roughly the same magnitude as in the cooler stars (Figure 3), and 
would not affect the discussion on Li abundance dispersion that follows in Section 6.
In the temperature and gravity range of our sample, non-LTE corrections to the Li abundance are 
estimated to be small (Carlsson et al. 1994); non-LTE corrections would lower the LTE 
abundances by 0.019 at the cool end of our sample and by 0.009 dex at the hot end.  Again, not incorporating
these relatively small corrections would not affect our discussion on the dispersion in Li abundance
that follows. 

Li abundances for the single-lined binaries in Table 1 and the non-members and
double stars in Table 2 were determined as for the single stars.  
The Li abundances of non-members should have the same accuracy as the remainder of
our sample and the Li abundance errors on the doubles is unknown.
However if cool companion lowers the temperature estimated for single-lined and 
double-lined binaries, those Li abundances will be proportionately lowered.  
In addition, double-lined binaries may have weaker 
Fe I and Li I lines due to continuum dilution. 
We have therefore marked the Li abundances of the SB2s as uncertain in Table 2.  If anything,
the Li abundances of these stars are likely to be larger than our estimates.
A stronger Li I line may result if a neighboring feature
line from the companion falls on the Li I line but we have 
measured the wavelength separation of the two components and are certain that 
the Li I feature is not contaminated in any of our SB2s.

\subsection{Iron abundance}
                                                                                                                             
Iron abundances were determined for 25 cluster members (Table 3).
 The typical measurement uncertainty in the EQW of a 40-60 m\AA\ Fe I line is $\pm$5 m\AA. 
An error of $\pm$100K in $T_{\rm eff}$, $\pm$0.25 dex in $\log g$, 
$\pm$0.1 kms$^{-1}$ in $\xi_t$ and $\pm$5 m\AA\  in Fe I EQW
results in Fe abundance errors of $\pm$0.07, $\pm$0.02, $\pm$0.02 and $\pm$ 0.05 respectively.   When these
uncertainties are combined, the resulting error in the
Fe abundance is $\pm$ 0.09.

The mean Fe abundance is 7.40$\pm0.08$ dex where this standard error is 
comparable to the estimate of the precision of a single determination. 
There may be a slight decrease in the derived Fe abundance with
decreasing temperature; stars with $T_{\rm eff} > 5500$ K give a mean  that
is 0.09 dex higher than stars with $T_{\rm eff} < 5500$ K. A similar
suggestion of a temperature dependence was made by BLS.
Since our Fe abundance is based on astrophysical $gf$-values for Fe\,{\sc i}
and Fe\,{\sc ii} lines and is derived using the solar abundance of 
$\log$N(Fe) = 7.50, the mean Fe abundance may be quoted as [Fe/H] = $-0.10$, with
[Fe/H] = $-0.04$ for $T_{\rm eff} > 5500$ K and [Fe/H] = $-0.13$ for stars 
with $T_{\rm eff} < 5500$ K. 
BLS obtained a mean Fe abundance about 0.13 dex higher
with astrophysical $gf$-values calculated from the empirical
Holweger-M\"{u}ller model (1974) and a microturbulence of 1.2 km s$^{-1}$.
Our result is in good agreement with Boesgaard \& Friel's (1990)
spectroscopic determination by an
essentially equivalent technique including the use of the Kurucz 
grid, though with fewer (15) Fe\,{\sc i} lines. They obtained
[Fe/H]$ = -0.054\pm0.046$ from six stars
with $T_{\rm eff}$ from 6415 K to 7285 K; our result from four stars
hotter than 6000 K is [Fe/H] $ = -0.04\pm0.08$. 

Also listed in Table 3 are the Fe abundance of the non-members, visual doubles and binaries 
from Table 2 for which spectroscopic
analysis was possible.  The mean Fe abundance of the four non-members is [Fe/H]=$-$0.13$\pm$0.15,
of the two double stars is [Fe/H] = $-$0.07$\pm$0.04, and of the six double-lined spectroscopic 
binaries is [Fe/H]=$-$0.14$\pm$0.21.
The mean Fe abundances are not very
different from that of the cluster members; the standard errors are slightly larger than for the 
cluster mean.

\begin{table*}
\caption{Iron abundances}
\begin{tabular}{cccccl}
\hline
Star &   & Model &   & {\it [Fe/H]} & Notes$^a$ \\
\# & $T_{\rm eff}$ & $\log g$ & $\xi_t$ &
& \\
  & (K) & cm s$^{-2}$ & km s$^{-1}$ &   &\\
\hline\hline 
Cluster Members \\
  56 & 5600 & 4.5 & 0.8 & + 0.06 & \\
 174 & 5000 & 4.3 & 2.0 & $-$ 0.25 & \\
 299 & 6200 & 4.5 & 1.3 & 0.00 & \\
 334 & 6400 & 4.5 & 1.7 & $-$ 0.18 & \\
 767 & 6100 & 4.5 & 1.3 & + 0.03 & \\
1086 & 5900 & 4.3 & 1.4 & $-$ 0.12 & \\
1185 & 6000 & 4.5 & 1.2 & $-$ 0.03 & SB \\
1514 & 5400 & 4.3 & 1.0 & $-$ 0.18 & \\
1525 & 5300 & 4.3 & 1.6 & $-$ 0.15 & \\
1528 & 4900 & 4.3 & 1.5 & $-$ 0.03 & \\
1537 & 5200 & 4.5 & 1.7 & $-$ 0.08 & \\
1538 & 5700 & 4.5 & 1.5 & $-$ 0.03 & \\
1565 & 4800 & 4.5 & 0.9 & $-$ 0.02 & \\
1570 & 5300 & 4.3 & 1.9 & $-$ 0.07 & \\
1572 & 5100 & 4.5 & 1.4 & $-$ 0.05 & \\
1575 & 4900 & 3.8 & 2.2 & $-$ 0.26 & SB \\
1578 & 5200 & 4.3 & 2.0 & $-$ 0.06 & \\
1604 & 5900 & 3.8 & 1.1 & $-$ 0.09 & \\
1606 & 4800 & 4.0 & 2.2 & $-$ 0.14 & \\
1610 & 5200 & 4.3 & 1.5 & $-$ 0.15 & \\
1621 & 5500 & 4.3 & 1.3 & $-$ 0.04 & \\
1669 & 4800 & 4.0 & 1.6 & $-$ 0.19 & \\
1697 & 5000 & 4.3 & 1.7 & $-$ 0.14 & \\
1731 & 4500 & 4.5 & 0.8 & $-$ 0.13 & \\
1735 & 4900 & 4.3 & 2.2 & $-$ 0.13 & \\
Non-members and Binaries\\
 143 & 5700 & 4.0 & 1.0 & $-$ 0.19 & NM, SB1O \\
 573 & 6600 & 4.0 & 0.6 & $-$ 0.23 & NM, SB \\
 848 & 6500 & 4.5 & 1.3 & $-$ 0.10 & SB2O \\
1100 & 5800 & 4.5 & 0.8 & + 0.09 & NM \\
1181 & 5700 & 4.0 & 1.1 & $-$ 0.19 & NM \\
1234 & 6000 & 4.5 & 1.6 & + 0.18 & SB2 \\
1538 & 5700 & 4.5 & 1.5 & $-$ 0.04 & Double \\
1541 & 5400 & 4.3 & 1.5 & $-$ 0.10 & Double \\
1602 & 5600 & 4.3 & 1.0 & $-$ 0.26 & SB2 \\
1625 & 5700 & 4.3 & 1.8 & $-$ 0.33 & SB2 \\
1656 & 5600 & 4.3 & 0.8 & $-$ 0.33 & SB2 \\
1713 & 5500 & 4.3 & 0.7 & + 0.02 & SB2 \\
\hline
\end{tabular}

\noindent
$^a$ Classifications as in Tables 1 and 2. Here NM denotes a non-member.

\end{table*}

\section{The Li abundance vs. Temperature relation}
                                                                                                                             
\noindent
The general nature of the relation between lithium abundance and effective temperature
was discussed previously by BLS and Randich et al. (1998).

In Figure 3, we show the Li vs $T_{\rm eff}(V-K)$ relation for the 70 stars in
Table 1 where the symbol's size  reflects {\it v} sin {\it i} as depicted in the legend
on the figure. The shaded region at the bottom of the figure displays the effect of a
correction to $T_{\rm eff}$ of 200 K across the range from 6400--4500 K.

\begin{figure}
\centering
\includegraphics[height=9cm,width=12cm]{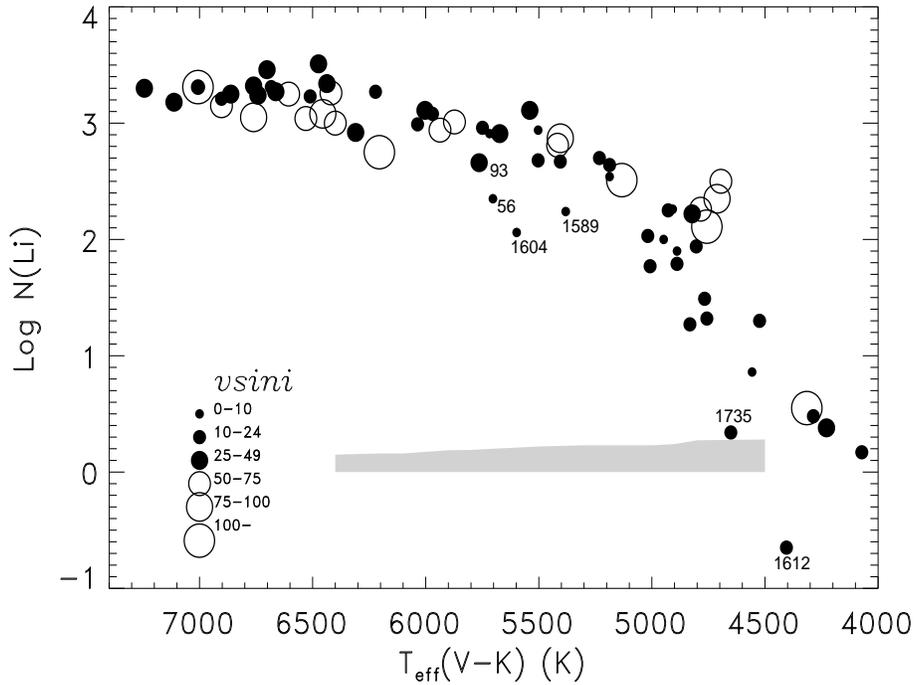}
\caption{The effective temperature vs lithium abundance relation for $\alpha$ Per.  The
projected rotational velocity ({\it v} sin {\it i}) of the stars is represented as in 
the legend. The shaded strip at the bottom of the figure shows the Li
abundance spread resulting from an effective temperature uncertainty of 200 K.
Several stars are labelled by the membership number for easy reference.}
\end{figure}

\subsection{The Hot Stars}
                                                                                                                             
\begin{table*}
\caption{Mean lithium abundances and dispersions in temperature bins.}
\begin{tabular}{ccc}
\hline
$T_{\rm eff}$ range & Mean & $N_{stars}$ \\
(K) & $\log$ N(Li) &  \\
\hline\hline 
Hot Stars\\
$>$ 7000 & 3.28 $\pm$ 0.06 & 4 \\
6750-7000 & 3.20 $\pm$ 0.10 & 5 \\
6500-6750 & 3.25 $\pm$ 0.12 & 7 \\
6250-6500 & 3.19 $\pm$ 0.22 & 6 \\
Middle Third \\
6000-6250 & 3.09 $\pm$ 0.13 & 4 \\
5750-6000 & 2.97 $\pm$ 0.23 & 3 \\
5500-5750 & 2.72 $\pm$ 0.34 & 9 \\
\hline
\end{tabular}
\end{table*}
                                                                                                                             
The Li vs $T_{\rm eff}(V-K)$ relation asymptotically approaches a constant
abundance at the high temperature end. In analysing this approach, we calculate mean
abundances in four temperature bins for stars hotter than 6250 K (Table 4). 
Each bin has roughly the same number of stars and the mean Li abundance 
in the four bins is essentially the same within the errors. 
In the three hottest bins, the dispersion within each bin can 
be easily accounted for by a combination of temperature errors ($\pm 200$K
corresponds to
$\pm 0.14$ dex), S/N of the spectra, and small reddening variations.
Whether the slightly larger dispersion is the coolest of the four bins is
significant, cannot be determined from our relatively small sample size.
Within our errors, there appears to be no dispersion in the Li distribution 
or significant change in mean Li value of these hottest stars.

Within these stars, there is an indication from Figure 3 that the most massive rapidly rotating stars 
(at $T_{\rm eff}(V-K)$ $> 6400$ K) have a slightly lower
Li abundance than the slow rotators. Taken as a whole, the results in Figure 3 may 
suggest that the rapidly rotating stars provide a relation with a shallower slope 
for $T_{\rm eff} > 5500$ K than the slowly rotating stars.
                                                                                              
The Li abundance in the hottest stars is equal to the meteoritic value
($\log$N(Li) = 3.25$\pm$0.06 according to Grevesse et al. 2007). It is this
value that has often been taken as a fair representation of the initial
value for young open clusters like $\alpha$ Per. Some authors also
quote a very similar  Li abundance derived from T Tauri stars (see, for example,
Magazzu et al.
1992 and Mart\'{\i}n et al. 1994). These two data points suggest that the local
value of the Galactic Li abundance has changed little in the last 4.5 Gyrs.

\subsection{The Middle Third}

Three temperature bins define the middle third of the sample between 5500 K and 
6250 K (Table 4).  The mean Li trend declines by 0.5 dex in this range.  
The dispersion in lithium appears to be larger 
than can be accounted for by the uncertainties in the stellar parameters and the S/N of our 
spectra.  In order to understand this dispersion, we examined
four cluster members, \#1589, \#1604, \#56 and \#93, 
with $T_{\rm eff}$ between 5400 K and 5750 K.  These appear to 
be outliers to what would otherwise be a fairly narrow mean Li trend similar to that
seen in the hotter stars; in the
absence of these stars, the decline in the mean Li trend between 6250 K and 5500 K would
be 0.3 dex and the dispersion in the coolest bin be only $\pm0.13$. 

It is worth noting that these four stars have much lower Li EQWs compared to other stars 
in the same temperature range. 
We therefore begin by examining three possible
explanations for these outliers: (i) their assigned effective
temperature is in error,
(ii) the stars are non-members
that have experienced normal Li depletion for their age, 
(iii) the dispersion in Li is not real but a reflection of differences in 
chromospheric activity levels which affect the formation of the Li I line and
thereby the the equivalent widths of line, and
(iv) an unusual amount of Li
depletion has occurred in these cluster members.
We comment on each of these in turn.

Errors in the estimated temperatures appear to be the least likely cause of 
the outlier stars.  An increase in effective temperature would increase the 
estimated Li abundance of the outlier, but as the mean Li trend increases with
increasing temperature, the required temperature increase is larger than
that indicated simply by the temperature difference between the outlier abundance and 
the mean trend
at that temperature.  Consider the case of \#1589 which is about 0.4 dex below
the mean relation.  A 500 K increase in effective temperature eliminates this
deficit but at the new temperature of 5900 K the star remains about 0.3 dex below 
the mean relation.  A similar problem arises if the effective temperature is lowered.
The temperature change required to meet the mean Li trend is even larger in \#56
and it cannot be reconciled with the errors we have
derived for our estimated temperatures.
For example, the estimated temperatures of \#56, $T_{\rm eff}$(spec) = 5600 K and 
$T_{\rm eff}(V-K)$ = 5703 K, are in good agreement and we see no reason to consider them
to be in error by 700 K or larger.
The other outliers would require similar and unacceptably large increases or 
decreases in temperature to put them on the mean Li trend.  
An added constraint against such a large increase in temperature is the measured Fe 
abundance.  The 700 K increase in effective temperature required for \#56 would 
increase [Fe/H] from the
measured value of +0.06 to an extraordinarily high value of +0.42. 
Similarly, \#1604 has 
a measured metallicity of [Fe/H]=$-$0.09, consistent with the cluster mean, and the
even larger increase in effective temperature increase would result in an unbelievable
Fe abundance.
An explanation for the quartet in terms of an error in their effective
temperatures is therefore not credible. 

Possibly, these outliers are not in fact cluster
members. 
In the case of \#56, Prosser (1992) assigned it a questionable status as a member 
on the basis of its radial velocity but 
full status on the basis of proper motion. 
If \#56 is an interloper star with
normal Li depletion for its age, it should be roughly the age of the Hyades cluster
(600 Myr) (Thorburn et al. 1993, 
Boesgaard \& Tripicco 1986, Boesgaard \& Budge 1988). Similarly, at roughly the same temperature, \#1604
with a slightly lower Li abundance, would be a somewhat older star.  
However, neither \#56 nor \#1604 could be as old as NGC 752 (2.4 Gyr) or M67 (4.5 Gyr) 
because by that age solar-temperature stars have Li abundances of logN(Li)= 1.5 or 
lower (see Balachandran 1995 and references therein). 
The likelihood of 
Hyades-age interlopers in the field of view of the 
$\alpha$ Per cluster and at the distance of the $\alpha$ Per cluster is small. 
Although Prosser (1992) assigned cluster membership to \#93 without a radial velocity 
measurement, the moderate rotational velocity of the star ({\it v} sin {\it i} = 25 km s$^{-1}$)
increases the likelihood that it is a young star and therefore a cluster member;  
G stars are observed to have spun down by the age of the Pleiades (Stauffer et al. 1984).
As noted in Table 1, Mermilliod et al. (2008) questioned the cluster membership of \#1589.
The relatively high Li abundance (log N(Li)=2.24) relative to field stars at the 
same temperature suggests that the star is young;
the abundance of Li is of order log N(Li)=1.0 at 5300 K even in a cluster as young as the Hyades.
Therefore, even if the quartet are rejected as members on this flimsy evidence, it is obviously
no simple matter to account for their Li abundance as field stars.  

The effect of chromospheric activity, in particular surface inhomogenieties in the form of spots 
and plages, on the formation of the Li I line, and the subsequent effect on the equivalent width of
the line has been the focus of several studies (Randich 2001; H\"{u}nsch et al. 2004, 
King \& Schuler 2004, Xiong \& Deng 2006, King et al. 2010).
Typically the K I resonance line, that is formed in the same part of the atmosphere as Li I, is measured 
for comparison.  Although a spread in K I equivalent widths has been observed in stars of the same
temperature in the Pleiades (Jeffries 1999)
and IC 2602 (Randich 2001), the authors state that while there is a need to understand K I differences, 
there is no conclusive evidence that the 
spread in Li abundances in these young clusters can be attributed to differences in chromospheric
activity alone. In a recent study of the high resolution spectra of 17 cool
Pleiades dwarfs, King et al. (2010) found that the Li I line strengths had a larger scatter 
than the K I $\lambda$ 7699 \AA\ line strengths. They concluded that there must be a true
abundance component to the Pleiades Li dispersion and suggested that it may be due to
differences in pre-MS Li burning caused by the effects of surface activity on stellar 
structure.
Here we add a few nuggets to that discussion.
Our KPNO spectra contain the 7699 \AA\ K I resonance line at the edge of one of the echelle orders.  
We were able to measure this K I feature in 14 stars.  The data show the expected increase in EQW with 
decreasing temperature but the sparsity of the data preclude a detailed analysis.  
In addition to the two outliers, \#1604 and \#1589, we were able to measure the K I EQWs of 
two normal stars at the same temperature \#1086 and \#1185.  The data are shown in  Table 5.
There are two findings of relevance.  First, while the Li I EQW of \#1604 is a factor of three smaller than 
that of \#1086 and \#1185, the K I EQWs of all three stars are within about 15 percent of each other.
Second, comparing the K I lines in the two stars with low Li, \#1589 and \#1604, we find the ratio of their 
KI EQWs is 1.6, perhaps reflecting the lower $T_{\rm eff}(V-K)$ of \#1589.  The Li I EQW ratio of the two stars is 1.7 and 
mirrors the K I EQW ratio.  Thus, we are able to discern no reason to attribute the low Li abundances
to \#1589 and \#1604 to the effects of chromospheric activity.

Since a convincing case cannot be made for either explanation (i), (ii), or (iii), the so-called 
outliers must be accepted as cluster members with an above-average depletion of lithium. 
With their inclusion as members, we conclude that a dispersion in lithium is clearly present
between 5500 K and 5750 K and, as will be discussed in the next sub-section, this
dispersion persists at cooler temperatures.

\begin{table*}
\caption{Comparison of K I and Li I data}
\begin{tabular}{cccccc}
\hline
Star \#  &  {\it $T_{\rm eff}(V-K)$} (K) & {\it $T_{\rm eff}(spec)$} (K) & $W_\lambda$(K I) (m\AA) &  
$W_\lambda$(Li I) (m\AA) & $\log$ N(Li) \\
Warm Stars\\
1086 & 5750 & 5900 & 192 & 151 & 2.96 \\
1185 & 5718 & 6000 & 194 & 132 & 2.91 \\
1589 & 5381 & 5900 & 350 & 77 & 2.24 \\
1604 & 5600 & 5900 & 221 & 45 & 2.06 \\
Cool Stars \\
174 & 4928 & 5000 & 308 & 196 & 2.25 \\
1697 & 4767 & 5000 & 391 & 78 & 1.49 \\
\hline
\end{tabular}
\end{table*}

\subsection{The Cool Stars}

Stars cooler than about 5500 K appear to fall in a widening band of declining lithium with 
decreasing temperature (Figure 3). There
seems to be a lower envelope, defined by low {\it v} sin {\it i} stars, running from
a Li abundance of about  
$\log$ N(Li) = 2.5 at 5500 K to $\log$ N(Li) =  $-$0.4  at 4500 K, and an upper envelope
running
through high {\it v} sin {\it i} stars K with a Li abundance of around $\log$ N(Li) = 
3.0 at 5500 K and then falling to a Li abundance of $\log$ N(Li) = 0.0 at 4200 K. 
The width of the band at
temperatures less than 4700 K is about 1.5 dex, a width resembling that for the
Pleiades, a cluster about 20 to 30 Myr older than $\alpha$ Per (Soderblom et al. 1993; Sestito
\& Randich 2005)).
Xiong \& Deng (2005) in a discussion on the Li abundances provided by the BLS sample
drew attention to a scatter appearing around the colour index (V$-$I$_c$) = 1.03 or
about 4700 K. This was about the cool end of BLS's sample. Our expanded sample shows that 
the scatter begins at a somewhat warmer temperature around 5600 K and  extends to cooler 
temperatures.

The Li distribution band is sketched in Figure 4.  Accepting the outliers discussed in
the previous section as members, the Li distribution band in the cooler stars can be extended
to warmer temperatures with the upper and lower bands asymptoting to the Li plateau value 
at 7000 K.  The dispersion in Li may begin at 6500 K, though additional stars are required to 
define this spread, and gradually widen in the cooler stars.

\begin{figure}
\centering
\includegraphics[height=9cm,width=12cm]{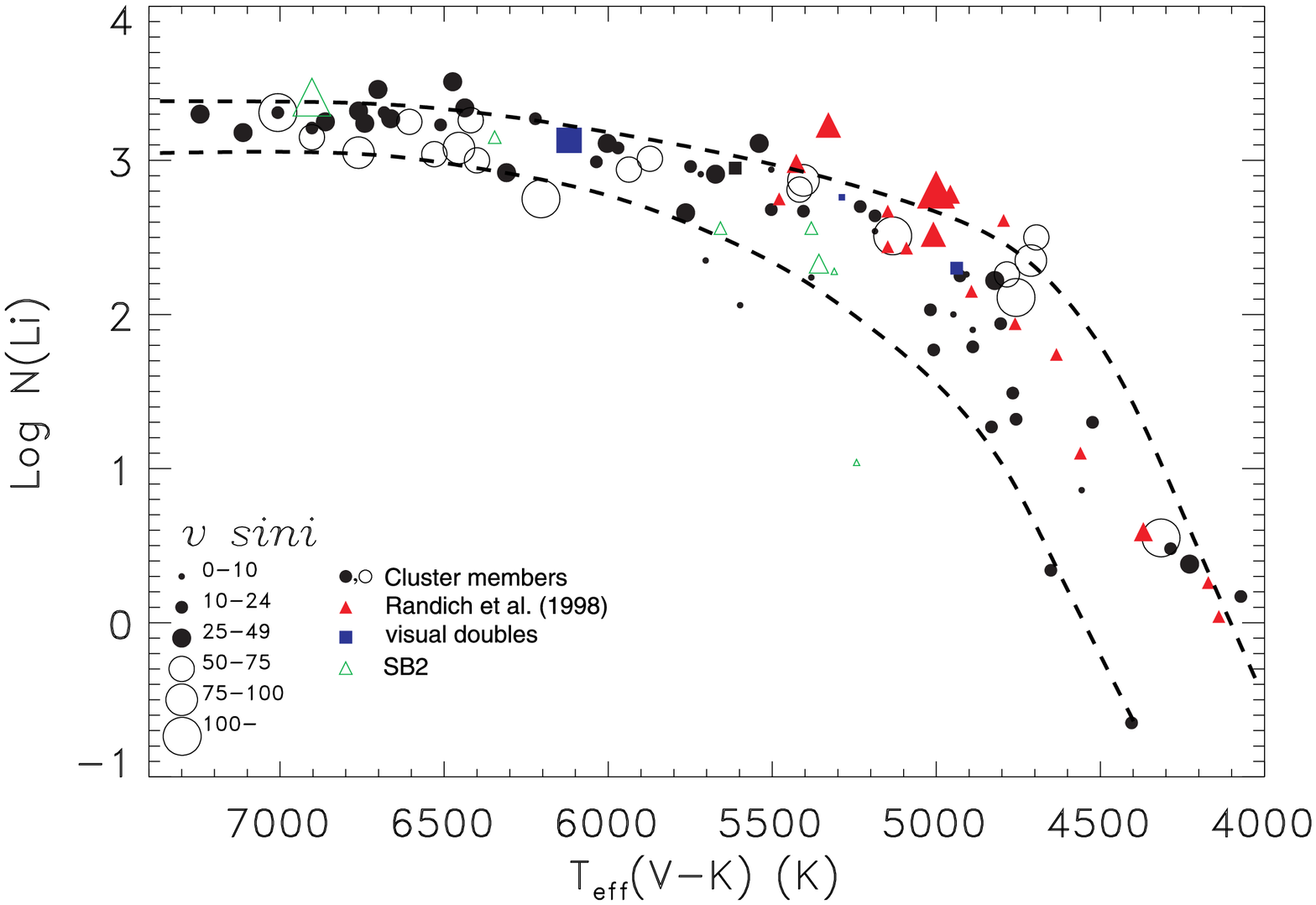}
\caption{The lithium abundance vs effective temperature relation for $\alpha$ Per
(as in Figure 3).  Red triangles are data added from Randich et al. (1998) with larger
symbols for faster rotators.  Blue squares and green triangles are visual doubles
and spectroscopic binaries from our sample. Suggested upper and lower envelopes to
the relation are indicated as dashed curves.}
\end{figure}
                                                                                                                             
Randich et al. (1998) provided Li abundances for 18  X-ray selected members of the
cluster. An additional five stars were analysed but declared to be non-members.
The spectra were at a resolution of 1\AA\ but lines blended with the Li\,{\sc i}
doublet were taken into account in the analysis.  
The adopted $T_{\rm eff}$ scale is
in good agreement
with ours. A comparison with their and our temperatures for the BLS sample indicates
a mean difference (Us$-$Them) of only 6$\pm67$K when two wildly discrepant stars are
excluded.  This suggests that we may add these X-ray selected stars to our sample.
Furthermore, we note that one star - \#1601 (AP101) - is a common star: we find
$\log$ N(Li) = 0.48 and
Randich et al. give 0.68, an unimportant difference given the  spread at
the 4300 K
temperature of the star.
In Figure 4,  abundances from Randich et al. are included along with ours and
lines drawn to represent the possible upper and lower envelopes to the Li abundance 
variation with effective temperature.
These additional stars at $T_{\rm eff} < 5000$ K tend to populate the
upper half of the band between our  suggested upper and lower
envelopes.
Unfortunately, the new points provide few high {\it v} sin {\it i} objects.

Scatter at temperatures below about 5500 K cannot
be attributed to the standard sources of uncertainty (incorrect effective
temperature,  uncertainties in measuring the 6707 \AA\ feature, contamination
of the sample by nonmembers, etc.).
Several ideas have been suggested linking the Li
abundance scatter at least in part to the failure of classical model atmospheres (as
used here)
to represent the real atmospheres of these young late-type dwarfs.
In Section 6.2 we discussed our K I data for four stars around 5900 K. Our data include 
K I equivalent widths for two additional stars at 5000 K, \#174 and \#1697, 
both bonafide cluster members (Table 5).   These data also do not provide any support 
for a link between high Li abundance and chromospheric activity.  Rather 
the larger K I EQW corresponds to the star with the smaller Li abundance.

\subsection {Doubles, Binaries and Non-members}

In Figure 4, the stars designated as double stars and double-lined spectroscopic 
binaries in Table 2 are shown by symbols of different colors as indicated in 
the legend accompanying the figure.

The four double stars with separations of 0.5 arc seconds or less lie well within the
Li distribution band of the normal stars.  The spectra of these stars appear to not have been
significantly contaminated by the presence of the nearby star and future analyses may simply include 
them as member stars.  

With the exception of \#1713, the double-lined spectroscopic binaries (green triangles) also lie well 
within the 
Li band of the cluster members.   Their measured Li abundances may be regarded as lower limits to the 
value that would be obtained if the continuum contamination of the secondary was properly accouted for.
Any further interpretation of their abundances would require a 
rigorous analysis that takes into account the continuum  and line spectrum of both stars.

Surprisingly three of the four stars identified as non-members by Mermilliod et al. (2008): \#143,
\#1181 and \#1100, have Li abundances that are entirely compatible with the mean cluster
trend.  The star \#1181 is coincident with \#588, a rapidly rotating cluster member, 
\#143 lies on the lower Li envelope of the cluster, and \#1100 is in the proximity of \#1604, the outlier 
star that we found no reason to exclude from the sample.   The three stars were deemed to be non-members by
Mermilliod et al. (2008) on the basis of their radial velocity and proper motion alone; all three
lie on the cluster's color-magnitude diagram and are therefore at the distance of the cluster.
This combination of facts makes the three stars rather enigmatic.  
The relatively large lithium abundances of these stars compared to field stars would make them not 
much older than a few hundred Myr;  the likelihood of relatively young interloper stars in the field of 
view of the cluster and at the distance of the cluster must be rather small.  

The two remaining stars identified as non-members, \#573  and \#407, may be field stars.  The former appears to
lie in the region of the Li-dip and the latter has a low enough Li abundance to be consistent with field star
values (Chen et al. 2001).  We note, however, that \#407 has a rather large rotational velocity 
({\it v} sin {\it i} = 28 km s$^{-1}$),
which is unusual in a field G star.  In summary, all of the five stars categorized as non-members 
may warrant closer
scrutiny.

\section{Concluding remarks}

In the Introduction, we referred to the powerful role that
is played by open clusters in placing observational
constraints on lithium depletion in pre-main sequence (PMS) and
main sequence (MS) stars. Perhaps, the principal outstanding
questions about lithium depletion concern the onset of the
depletion  and the star-to-star spread in (apparent)
lithium abundances at low masses. (There remains too the incompletely
understood Li-dip in warm older stars.)
In order to determine when lithium depletion  at low masses
develops, and how it evolves with time,  depletion of 
Li must be traced from the earliest PMS phases to the 
age of the $\alpha$ Per cluster and beyond.

PMS lithium depletion is
now mappable by looking at the very youngest of clusters and associations;
the stars are faint but accessible with large telescopes. 
In addition to the uncertainty of defining membership in clusters, associations
and moving groups, 
there are several problems associated with interpreting their 
Li abundance trends.
First, 
since young PMS stars of different masses
tend to  lie in the same temperature range 
between 3000 K to 4000 K as they evolve
down the HR diagram, and older PMS stars rapidly increase their 
temperature as they
evolve towards the main sequence, the temperature of the PMS 
star is not  a sufficient indication of its mass and
determination of stellar mass requires  the use of theoretical evolutionary
tracks which continue to differ from author to author. This difficulty may be
offset partly by the result that theoretical prediction of a cluster's  age
from the location of the lithium depletion boundary (LDB) is not very
dependent on which set of PMS evolutionary tracks is chosen (Jeffries \&
Oliveira 2005). 
Second,
 because the young PMS stars are  cool, analysis of their spectra 
is complicated by molecular features and the derived Li abundance has a
larger uncertainty than in warmer main sequence stars.

For these reasons and because the available data are limited 
with respect both  to the number of
young clusters, associations, and moving groups and to the number of
stars per cluster, we defer a detailed search for the onset of lithium
depletion at low masses.  
It is worth noting that an interesting set of Li data in a range of young clusters and 
associations 
has been accumulated ({\it e.g.} Sestito \& Randich 2005, Mentuch et al. 2008).
In their Table 1, Sestito \& Randich (2005) list the
'classical' ages of clusters, i.e., those determined from isochrone fitting. For four young
clusters in their sample, IC 2391, NGC 2547, $\alpha$ Per and the Pleiades, new independent
estimates of the ages have been obtained based on the position of LDB (Stauffer et al. 1998, 1999;
Barrado y Navascues et al. 2004, Jeffries \& Oliviera 2005) which are, in general, higher 
than the classical ages. Although the LDB technique is less model dependent than MS fitting,
Sestito \& Randich chose to adopt  the classical ages for uniformity through the entire 
sample. 
In our discussion of clusters chosen from Sestito \& Randich
(2005), including $\alpha$ Per and the Pleiades, we have adopted these same classical ages.

On the other hand, the ages of the young associations studied by Mentuch et al. (2008), that we
compare the results of $\alpha$ Per and the Pleiades to, are derived by comparing the 
dependence of Li abundance on temperature with isochrones from pre-MS evolutionary tracks. 
Mentuch et al. state that these ages are consistent with the earlier estimates based on 
isochrone fitting or other methods.   We will not analyze the strengths and weaknesses of
cluster age determinations in our discussions. Rather, we note that the crucial point
is that the ordering of the clusters according to age is robust. When comparing the results
of Li scatter in these young associations with those of clusters from Sestito \& Randich
(2005) with classical ages, the chronological order of ages is not disturbed even if we adopt
the higher LDB ages for the 4 young clusters, namely, NGC 2547, IC 2391, $\alpha$ Per and the
Pleiades. It is therefore worth examining the star samples of these associations with those
of $\alpha$ Per and other clusters.
 
Sestito \& Randich list NGC 2264 at 5 Myr as their youngest
cluster with the survey of Li abundances from Soderblom et al. (1999).
The Li abundance for the warmer stars in NGC 2264 is about 3.2, a value
consistent with our result for the hotter stars in $\alpha$ Per
and also with the canonical value for an initial Li abundance for young stars.
Stars between 0.5 and 1.0 M$_\odot$ in NGC 2264
show only a mild (0.4 dex) Li dispersion.  
Given measurement and analyses uncertainties, one
may conclude that PMS depletion has possibly
 not begun in these young stars.  Mild PMS depletion may be
seen in the 12 Myr old $\eta$Cha cluster and TW Hydrae association (Mentuch
et al. 2008).  By 20 Myr, the $\beta$ Pic moving group and by 27 Myr the
Tucanae-Horologium
association show nearly a 3.0 dex range in Li abundance (see Figure 8 in Mentuch et
al. 2008).
However, in the absence of reliable mass determinations, the presence or absence of
an abundance
disperson at a particular mass cannot be deciphered.

A clearer view of Li dispersion at a particular mass may be obtained once the cluster
is on the main sequence.
Sestito \& Randich (2005)  list IC 2602, IC 2391, IC 4665, and NGC
2547 as main sequence clusters younger than $\alpha$ Per.
Impression of a smaller star-to-star scatter in Li abundances in clusters
younger than $\alpha$ Per is conveyed by results for IC 2602 with
an age of 30 Myr (Randich et al. 1997, 2001). Randich et al. (2001)
define a regression curve to represent Li abundances from
3900 K to 6900 K. This curve is above the upper envelope in Figure 4 for
$T_{\rm eff} < 4400$ K and coincident with it for higher temperatures.
To effect a fair comparison, a correction would need to be made for the
mass-dependent temperature change between an age of 30 Myr and 50 Myr.
Little additional Li depletion is predicted in this interval.  The point of
interest here is that the scatter about the regression curve is at most
$\pm0.5$dex, often much less, and less than exhibited in Figure 3. Indeed, most
points in the equivalent plot to Figure 3 (Randich et al.'s (2001) Figure 4)
touch the regression curve with their error bars. 
A similar conclusion applies to IC 2391, also
30 Myrs old, from inspection of the same Figure 4 which assembles
Li abundances from that paper and Stauffer et al. (1989b).
For IC 4665 at 35 Myr, the available Li abundances (Mart\'{\i}n \& Montes 1997;
Jeffries et al. 2009) are too few at low temperatures to define the Li abundance
trend with temperature and certainly not to detect a star-to-star variation.
For NGC 2547  also at 35 Myr, there is evidence of a variation
approaching that seen in Figure 3 (Jeffries et al. 2003) with a lower
envelope to the Li abundances resembling that of the upper envelope in
Figure 4.
One may speculate from these comparisons 
that the dispersion in Li at a given mass
develops and strengthens between 30 and 50 Myr, that is
 between the ages of IC 2391 and IC 2602 and the
age of $\alpha$ Per.  

 This is further corroborated by observations of the AB Doradus moving group by 
Mentuch et al. (2008) which has an age of 45 Myr, very similar to that of $\alpha$ Per.
With the additional caveats that the sample is small and there are membership issues in
defining a moving group, we compare our $\alpha$ Per sample with that of AB Dor.
The comparison is frustrated because
Mentuch et al. (2008) systematically find an abundance $\log$N(Li) $\simeq 3.8$
in their samples for stars that are unaffected by PMS depletion. This `initial' value  is
about 0.6 dex greater than our value for the hotter stars.
We have adopted the view that the high initial abundance is a consequence
of a systematic overestimate of the Li abundance but we have no basis for knowing if this
overestimate carries over to lower temperatures.  Between 5300 K and 4900 K, four AB Dor stars 
have Li abundances between 3.4 and 1.3; the range
is comparable to that seen in $\alpha$ Per and larger than that seen in the Sestio \& Randich (2005)
survey.  Below 4900 K, five AB Dor stars are coincident with
the upper envelope of $\alpha$ Per stars.  There may be some indication in this limited sample
that AB Dor exhibits a larger Li spread than the slightly younger clusters discussed
in Sestito \& Randich (2005) but very similar to what is seen in $\alpha$ Per.
 
For clusters older than $\alpha$ Per, we restrict comparison to the
well-sampled Pleiades (age of 70 Myr) where we have taken
Pleiades data from Soderblom et al. (1993, see also King et al. 2000).
Perhaps, a fairer comparison would be to take data for both
clusters from Sestito \& Randich (2005) who undertook a uniform
analysis of these and other open clusters. The mean relations and their
scatter are very similar but for two minor differences when compared
in the abundance-effective temperature plane; the evolution in $T_{\rm eff}$
over the 20 Myr age difference is very small and ignored. First, the $\alpha$
Per cool outliers -- \# 1612 and \#1735 -- have no counterparts
in the Pleiades. Second, and more prominently, the Pleiades has four stars with
undepleted lithium ($\log$N(Li) $\simeq 3.2$) at $T_{\rm eff} \simeq 5000$ K with no
counterparts in $\alpha$ Per where $\log$N(Li) $\simeq$ 2.5 at this temperature.
                                                                                                                             
In observed  clusters older than Pleiades, main sequence depletion
begins to reduce the Li abundances in the coolest stars noticeably. This is
certainly apparent for M34 with an age of 250 Myr where stars have been observed
down to about 4200 K (Jones et al. 1997). Here, the star-to-star scatter remains
similar to that of the Pleiades and $\alpha$ Per but the mean abundances
are smaller. By the age of the Hyades, only upper limits to the Li I equivalent
width are measurable in stars $\leq$ 5000 K (Soderblom et al. 1995).  
                                                                                  
In summary, the Li abundances for $\alpha$ Per  fit
the pattern provided by observations of clusters both younger and
older than it. 
The star-to-star spread appears to develop after about 20 Myr. 
The spread survives up to  250 Myr and  its
demise is hidden from observers as main sequence lithium
depletion removes any inequalities
in lithium abundance from observers' view.   

Inspection of Figures 3 and 4 shows a relative dearth of measurements at
temperatures lower than about 4700 K. Additional members of the
$\alpha$ Per cluster  are to be
found in Prosser (1992). Although expansion of the sample at lower
temperatures would be informative,  
perhaps the most useful benefit from  an enlarged sample, 
would be an application of the best techniques
of quantitative stellar spectroscopy to pairs of stars
with maximum and minimum Li abundance but similar observed properties
such as colour and rotation period.  If such a study discovers differences 
only for lithium, then atmospheric effects may truly be eliminated as the
cause of the Li dispersion.

\section*{Acknowledgments}
This research has made use of the WEBDA database maintained  at the Institute for
Astronomy of
the University of Vienna.
This publication makes use of data products from the Two Micron All Sky Survey, which is a joint project of the University of Massachusetts and the Infrared Processing and Analysis Center/California Institute of Technology, funded by the National Aeronautics and Space Administration and the National Science Foundation. We thank the referee for several constructive
comments on the manuscript.
SCB is pleased to acknowledge support from NSF grant AST-0407057.  On the eve of her departure from 25 years of
research in Astronomy, SCB would like to thank colleagues and collaborators who have made it
a rich and satisfying experience.
 DLL
wishes to thank the Robert A. Welch Foundation of Houston, Texas for
support through grant F-634.
The authors would like to thank the referee for helpful remarks that improved the manuscript.


\begin{thebibliography}{}
\bibitem[ ]{} Alonso, A., Arribas, S., \& Mart\'{\i}nez-Roger, C., 1996, A\&A, 313, 873
                                                                                                                          
\bibitem[ ]{} Andersen, J., Gustafsson., B., \& Lambert, D.L., 1984, A\&A, 136, 65


\bibitem[ ]{} Balachandran, S., 1995, ApJ, 446, 203
                                                                                                                          
\bibitem[ ]{} Balachandran, S., Lambert, D.L., \& Stauffer, J.R., 1988, ApJ, 333, 267
                                                                                                                          
\bibitem[ ]{} Balachandran, S., Lambert, D.L., \& Stauffer, J.R., 1996, ApJ, 470, 1243
                                                                                                                          
\bibitem[]{} Barrado y Navascu\'{e}s, D., Stauffer, J.R., \& Jayawardhana, R., 2004,
ApJ, 614, 386


\bibitem[ ]{} Boesgaard, A.M., \& Budge, K.G., 1988, ApJ, 332, 410
                                                                                                                          
\bibitem[ ]{} Boesgaard, A.M., Budge, K.G., \& Ramsay, M.E., 1988, ApJ, 327, 389
                                                                                                                          
\bibitem[ ]{} Boesgaard, A.M., \& Friel, E.D., 1990, ApJ, 351, 467
                                                                                                                          
\bibitem[ ]{} Boesgaard, A.M., \& Tripicco, M.J., 1986, ApJ, 302, L49
                                                                                                                          
\bibitem[]{}  Butler, R.P., Cohen, R.D., Duncan, D.K., \& Marcy, G.W., 1987, ApJ,
319,  L19
                                                                                                                          
\bibitem[ ]{} Carlsson, M., Rutten, R.J., Bruls, J.H.M.J., \& Shchukina, N.G., 1994,
A\&A,
288, 860
                                                                                                                          
\bibitem[ ]{} Carpenter, J.M., 2001, AJ, 121, 2851
                                                                                                                          
\bibitem[ ]{} Castelli, F., Gratton, R.G., \& Kurucz, R.L., 1997, A\&A, 318, 841
                                                                                                                          
\bibitem[ ]{} Castelli, F., \& Kurucz, R.L., 2006, A\&A, 454, 333


\bibitem[ ]{} Chen, Y.Q., Nissen, P.E., Benoni, T., \& Zhao, G., 2001, A\&A, 371, 943
                                                                                                                          
\bibitem[ ]{} Crawford, D.L., 1975, AJ, 80, 955


\bibitem[ ]{} Crawford, D.L., \& Barnes, J.V., 1974, AJ, 79, 687
                                                                                                                          
\bibitem[]{} Delbouille, L., Roland, G., \& Neven, L.\ 1990, Liege: Universite de Liege, 
Institut d'Astrophysique, 1990 


\bibitem[ ]{} Garc\'{i}a L\'{o}pez, R.J., Rebolo, R., \& Mart\'{\i}n, E.L., 1994,
ApJS, 90, 531
                                                                                                                          
\bibitem[]{}  Grevesse, N., Asplund, M., \& Sauval, A.J., 2007, SSRv, 130, 105
                                                                                                                          
\bibitem[ ]{} Heckmann, V.O., Dieckvoss, W., \& Kox, H., 1956, AN, 283, 109
                                                                                                                          
\bibitem[ ]{} Heckmann, V.O., \& L\"{u}beck, K., 1958, ZfAp, 45, 243
                                                                                                                          
\bibitem []{} Holweger, H., \& M\"{u}ller, E.A., 1974, Solar Phys., 35, 19


\bibitem []{} H\"{u}nsch, M., Randich, S., Hempel, M., Weidner, C., \& Schmitt, J.H.M.M., 2004,
A\&A, 418, 539


\bibitem []{} Jeffries, R.D., 1999, MNRAS, 309, 189
                                                                                                                          
\bibitem[]{}  Jeffries, R.D., \& Oliviera, J.M., 2005, MNRAS, 358, 13


\bibitem []{} Jeffries, R.D., Jackson, R.J., James, D.J., \& Cargile, P.A., 2009, MNRAS,
in press
                                                                                                                          
\bibitem[]{}  Jeffries, R.D., Oliviera, J.M., Barrado y Navascu\'{e}s, D.,  \&
Stauffer, J.R., 2003, MNRAS, 343, 1271
                                                                                                                          
\bibitem[]{} Jones, B.F., Fischer, D., Shetrone, M., \& Soderblom, D.R., 1997, AJ,
114, 352
                                                                                                                          
\bibitem[ ]{} King, J.R., \& Schuler, S.C., 2004, AJ, 128, 2898


\bibitem[ ]{} King, J.R., Krishnamurthi, A., \& Pinsonneault, M.H., 2000, AJ, 119, 859
                                                                                                                          
\bibitem[]{} King, J.R., Schuler, S.C., Hobbs, L.M., \& Pinsonneault, M.H., 2010, ApJ, 
710, 1610
                                                                                                                          
\bibitem[]{}  Magazzu, A., Rebolo, R., Pavlenko, Ya.V., 1992, ApJ, 392, 159
                                                                                                                          
\bibitem[ ]{} Makarov, V.V., 2006, AJ, 131, 2967
                                                                                                                          
\bibitem[]{} Mart\'{\i}n, E.L., \& Montes, D., 1997, A\&A, 318, 805
                                                                                                                          
\bibitem[]{}  Mart\'{\i}n, E.L., Rebolo, R., Magazzu, A., \& Pavlenko, Ya.V., 1994,
A\&A, 282, 578
                                                                                                                          
\bibitem []{} Mentuch, E., Brandeker, A., van Kerkwijk, M.N., Jayawardhana, R., \&
Hauschildt, P.H., 2008, ApJ, 689, 1127
                                                                                                                          
\bibitem[ ]{} Mermilliod, J.-C., Queloz, D., \& Mayor, M., 2008, A\&A, 488, 409
                                                                                                                          
                                                                                                                          
\bibitem[ ]{} Mitchell,R.I., 1960, ApJ, 132, 68
                                                                                                                          
                                                                                                                          
\bibitem[ ]{} Patience, J., Ghez, A.M., Reid, I.N., \& Matthews, K., 2002, AJ, 123, 1570
                                                                                                                          
\bibitem[ ]{} Pe\~{n}a, J.H., \& Sareyan, J.-P., 2006, RMxAA, 42, 179
                                                                                                                          
\bibitem[ ]{} Prosser, C.F., 1992, AJ, 103, 488
                                                                                                                          
\bibitem[]{} Randich, S., 2001, A\&A, 377, 512
                                                                                                                          
\bibitem[]{} Randich, S., Aharpour, N., Pallavicini, R., Prosser, C.F., \& Stauffer,
J.R., 1997, A\&A, 323, 86
                                                                                                                          
\bibitem[ ]{} Randich, S., Mart\'{i}n, E.L., Garc\'{\i}a L\'{o}pez, R.J., \&
Pallavicini, R., 1998,
A\&A, 333, 591
                                                                                                                          
\bibitem []{} Randich, S., Pallavicini, R., Meola, G., Stauffer, J.R., \&
Balachandran, S., 2001, A\&A, 372, 862
                                                                                                                          
\bibitem[ ]{} Sestito, P., \& Randich, S., 2005, A\&A 442, 615
                                                                                                                          
\bibitem[ ]{} Sneden, C., 1973, Ph.D. Thesis, University of Texas


\bibitem[]{}  Soderblom, D.R., Jones, B.F., Balachandran, S., Stauffer, J.R.,
Duncan, D.K., Fedele, S.B., \& Hudson, J.D., 1993, AJ, 106, 1059
                                                                                                                          
\bibitem[]{} Soderblom, D. R., Jones, B.F., Stauffer, J.R., \& Brian, C., 1995, AJ, 110, 729


\bibitem[]{}  Soderblom, D.R., King, J.R., Siess, L., Jones, B.F., \& Fischer, D., 1999,
AJ, 118, 1301


\bibitem []{} Stauffer, J.R., Hartmann, L.W.,  Soderblom, D.R., \& Burnham, N., 1984,
ApJ, 280, 202


\bibitem []{} Stauffer, J.R., Hartmann, L.W., \& Jones, B.F., 1985, ApJ, 289, 247


\bibitem []{} Stauffer, J.R., Hartmann, L.W., \& Jones, B.F., 1989a, ApJ, 346, 160
                                                                                                                          
\bibitem[]{} Stauffer, J.R., Hartmann, L.W., Jones, B.F., \& McNamara, B.R., 1989b, ApJ, 342, 285
                                                                                                                          
\bibitem[]{} Stauffer, J.R., Schultz, G., \& Kirkpatrick, J.D., 1998, ApJ, 499, L199


\bibitem[]{} Stauffer, J.R., Barrado y Navascu\'{e}s, D., Bouvier, J., et al., 1999, ApJ,
527, 219


\bibitem[ ]{} Stauffer, J.R., Jones, B.F., Backman, D., Hartmann, L.W., Barrado y
Navascu\'{e}s,
D., Pinsonneault, M., Terndrup, D.M., \& Muench, A.A., 2003, AJ, 126, 833


\bibitem []{} Thorburn, J.A., Hobbs, L.M., Deliyannis, C.P., \& Pinsonneault, M.H., 1993,
ApJ, 415, 150
                                                                                                                          
\bibitem[ ]{} Trullols, E., Rossell\'{o}, G., Jordi, C., \& Lahulla, F., 1989, A\&AS,
81, 47
                                                                                                                          
\bibitem[ ]{} Tull, R.G., MacQueen, P.J., Sneden, C., \& Lambert, D.L., 1995, PASP,
107, 251
                                                                                                                          
\bibitem[ ]{} Xiong, D.-R., \& Deng, L., 2005,ApJ, 622, 620


\bibitem []{} Xiong, D.-R., \& Deng, L., 2006, ChA\&A, 30, 24
                                                                                                                          
\bibitem[ ]{} Zapatero Osorio, M.R., Rebolo, R., Mart\'{\i}n, E.L., \& Garc\'{\i}a
L\'{o}pez, R.J., 1996,
ApJ, 469L, 53
                                                                                                                          
\end{thebibliography}
\end{document}